\documentclass[letter]{aa} 

\usepackage{graphicx}
\usepackage{txfonts}
\usepackage{siunitx}
\usepackage{easy-todo}
\usepackage{ulem}
\newcommand{\hcop}[0]{HCO$^+$}
\newcommand{\dcop}[0]{DCO$^+$}

\newcommand{\electr}[0]{$e^-$}

\begin{document}

   \title{Machine learning-accelerated chemistry modeling of protoplanetary disks}

   \author{Grigorii V. Smirnov-Pinchukov
          \inst{1}
          \and
          Tamara Molyarova
          \inst{2}
          \and
          Dmitry A. Semenov
          \inst{1, 3}
          \and
          Vitaly V. Akimkin
          \inst{2}
          \and
          Sierk~van~Terwisga
          \inst{1}
          \and
          Riccardo Francheschi
          \inst{1}
          \and
          Thomas Henning
          \inst{1}
          }

   \institute{Max Planck Institute for Astronomy,
              K{\"o}nigstuhl 17, D-69117 Heidelberg, Germany\\
              \email{smirnov@mpia.de}
        \and
            Institute of Astronomy, Russian Academy of Sciences,
            Pyatnitskaya str. 48, Moscow, 119017, Russia\\
         \and
             Department of Chemistry, Ludwig Maximilian University,
             Butenandtstr. 5-13, D-81377 Munich, Germany
             }

   \date{Received \today; accepted}


  \abstract
    {}
   {With the large amount of molecular emission data from (sub)millimeter observatories and incoming James Webb Space Telescope infrared spectroscopy,  access to fast forward models of the chemical composition of protoplanetary disks is of paramount importance.}
   {We used a thermo-chemical modeling code to generate a diverse population of protoplanetary disk models. We trained a K-nearest~neighbors (KNN) regressor to instantly predict the chemistry of other disk models.}
   {We show that it is possible to accurately reproduce
chemistry using just a small subset of physical conditions, thanks to correlations between the local physical conditions in adopted protoplanetary disk models. We discuss the uncertainties and limitations of this method.}
   {The proposed method can be used for Bayesian fitting of the line emission data to retrieve disk properties from observations. We present a pipeline for reproducing the same approach on other disk chemical model sets. }


   \maketitle
%

\section{Introduction}

Time-dependent gas-grain chemical kinetics codes are widely used in the astrochemical modeling of the ISM, protoplanetary disks, and even exoplanetary atmospheres. A typical computational time to calculate the chemical evolution of an isolated volume over $10^5-10^6$~years usually takes about 0.1--10 seconds on a single CPU core, depending on the complexity of the chemical network and variability of the physical conditions. More complex networks -- that include deuterium or carbon isotopologues \citep{2013ApJS..207...27A,2016ApJ...822...53Y} or that separately treat reactions on dust surface and in the bulk of icy mantles \citep{2009ApJ...691.1459V, 2013ApJ...765...60G} -- increase the number of reactions and slow down the calculations drastically. For a typical grid size of a 2D protoplanetary disk model of $100 \times 100$, a session of chemical kinetics modeling takes at least 15 minutes up to several days on a single CPU core. This is a reasonable timeframe for forward modeling but too long for any retrieval of disk physical parameters based on a fitting of molecular data. Recently, \citet{2022ApJ...927..203K}  proposed a new method for deriving the physical parameters of uniform gas clouds using Markov chain Monte Carlo sampling with realistic chemical and radiative transfer modeling and successfully applied it to the L1544 data. However, protoplanetary disks exhibit more complex structures with strong gradients of physical conditions and another approach is thus needed.

With the recent large programs on protoplanetary disk chemistry at the Atacama Large Millimeter/submillimeter Array (ALMA) (MAPS: \citet{2021ApJS..257....1O_MAPS1}; 2021.1.00128.L/AGE-PRO PI: Ke Zhang, in progress; see also \citet{2021ApJS..257....6G,2021ApJS..257....9I}) and Northern Extended Millimeter Array (NOEMA) (L19ME/PRODIGE, PI: P. Caselli, Th. Henning., in progress, see Semenov et al., in prep.), as well as large, partially spatially unresolved surveys of circumstellar disk populations \citep[ODISEA:][]{2019MNRAS.482..698C, 2016ApJ...828...46A}, the amount of available data for analysis expands rapidly. In addition, high angular- and frequency-resolution observations of molecular emission lines in individual disks are now common \citep[e.g.,][]{2020ApJ...890..142P,2020A&A...636A..65G,2021ApJ...911..150P}. Analyses of such data can be vastly improved if all the lines are taken into account simultaneously \citep{2020A&A...638A.110F, 2022A&A...658A.103H}. However, protoplanetary disk model retrieval in a Bayesian sense is still in its infancy.

Our goal is to build a function (an estimator) to return the reasonably accurate chemical composition using the smallest possible subset of local physical conditions characterizing the protoplanetary disk. Applications of machine learning techniques are emerging in all fields of astronomy and physics \citep[e.g.,][]{2015MNRAS.450.1441D, 2018RPPh...81g4001D, 2019RvMP...91d5002C,2020A&A...642A.171R, 2022A&A...662A.108A}, providing a robust, human-independent, and flexible way to find dependencies and correlations within a data set. The approach is also being applied to astrochemistry: among others, \citet{2021ApJ...917L...6L} searched for similarities between molecules to propose possible detectable molecules, \citet{2021arXiv210409516G} suggested a method to reduce the number of species for chemical kinetics, \citet{2021A&A...653A..76H} explored a large parameter space to provide time-dependent chemistry, and, most recently, \citet{2022arXiv220703906V} presented a way to predict binding energies of molecules on dust surfaces.

In this work, we present a solution to the problem of chemical model performance. We computed a grid of 540 thermo-chemical protoplanetary disk models, containing more than two million physical bins in various disk environments, but at fixed elemental and dust compositions and at the same age, which we describe in Section 2. We used a K-nearest~neighbors (KNN) machine learning (ML) algorithm \citep{NearestNeighbors} to create a robust interpolation between local physical conditions in the disk and the abundances of molecular and atomic species, described in detail in Section 3. Once the estimator is trained, the chemistry can be predicted in milliseconds per disk model, making it much faster than the next bottleneck, namely, the line radiative transfer. In Section 4, we demonstrate the performance and limitations of the method on a small set of chemical species. In Section 5, we summarize our findings.

\section{Thermo-chemical protoplanetary disk model grid}

Chemical kinetics models, such as ALCHEMIC \citep{2010A&A...522A..42S}, NAUTILUS \citep{2016MNRAS.459.3756R}, UCLCHEM \citep{2017AJ....154...38H}, and KROME \citep{2014MNRAS.439.2386G} can be very flexible and include dozens to tens of thousands of parameters. Some of these parameters describe local physical conditions, such as gas and dust density and temperature as well as the local ultraviolet (UV) radiation field. Others are global parameters, for instance, the ionization rate by radioactive nuclides or such details as the probability modificator of a gas particle sticking to a dust grain after a collision, which affect a large fraction of the considered chemical reaction network. The initial elemental or molecular composition of the matter make up another set of important input parameters. In addition, the chemical network itself contains an extensive data set of reaction rates, with only $\sim 30\%$ of the rate values known to an adequate level of accuracy.

Classical chemical kinetics codes can utilize any combination of these parameters, but most parameters are  fixed in real applications. Also, some parameters could correlate between various chemical calculations. For example, it is reasonable to expect that the low-density, high-temperature regions of protoplanetary disks lie in the disk atmosphere, where the chemistry is dominated by a limited set of photo- and gas-phase reactions. On the other hand, cold and dense disk regions are typical for the midplane, where photochemistry is much less important and gas-grain interactions and surface reactions play a major role. This simple observation has led to the idea of constructing simplified chemical models that allow for quick computations that are nonetheless feasible to establish the abundances of simple species such as CO, without the need for full chemical modeling \citep{WilliamsBest2014}. Unfortunately, such approaches cannot be easily generalized to predict disk chemical composition for a larger set of observed molecules or important coolants.

To create a reference data set of protoplanetary disks with known chemical structures, we used the ANDES astrochemical model of a 2D axisymmetric hydrostatic disk. It employs a chemical network based on the ALCHEMIC network \citep{2011ApJS..196...25S}, with deuterium-bearing molecules and deuterium fractionation included following \citet{2013ApJS..207...27A, 2014ApJ...784...39A}. The network describes 1247 species and 38347 reactions, including gas-phase and surface two-body reactions, adsorption and reactive desorption, photoreactions, and ionization or dissociation by X-rays, cosmic rays, and radioactive nuclides \citep{2013ApJ...766....8A,2017ApJ...849..130M,2018ApJ...866...46M}. The rates of surface reactions are adjusted to mimic the chemical inactivity of the bulk icy mantles; they are multiplied by a factor equal to the fraction of the upper layers in the total number of surface particles. Following \citet{2016A&A...595A..83E}, we adopted an icy molecular initial composition based on the abundances of prestellar cores \citep{2011ApJ...740..109O}. We ran the time-dependent chemical evolution for the typically assumed age of 1\,Myr \citep{1998A&A...338..995W, 2002A&A...386..622A}.

   \begin{table}

      \caption[]{ANDES input grid}
         \label{tab:andesinp}
         \centering
        \begin{tabular*}{\linewidth}{@{\extracolsep{\fill}}ll}
            \hline \hline
            \noalign{\smallskip}
            Parameter      &  Value \\
            \noalign{\smallskip}
            \hline
            \noalign{\smallskip}
            $M_{\rm star}$ & $0.2,\ 0.3,\ 0.4,\ 0.5, \ 0.8,\ 1.0\ M_\sun$ \\
            $M_{\rm disk}$ & $0.001,\ 0.003,\ 0.01,\ 0.03, \ 0.1 \ M_{\rm star}$ \\
            $R_{c}$ & $20,\ 50,\ 80,\ 100,\ 130,\ 170\ $au\\
            $L_X$ & $10^{29},\ 10^{30},\ 10^{31}\ $erg s$^{-1}$\\

            \noalign{\smallskip}
            \hline
            \noalign{\smallskip}
            $T_{\rm star}$, $R_{\rm star}$ & $f(M_{\rm star})$ \citet{2008ASPC..387..189Y}   \\
            UV excess & $B_\nu(T_{\rm eff}= 10000\ \mathrm{K})$\\
            Accretion rate & \num{1e-8} $ M_\sun \rm{yr}^{-1}$ \\
            Dust opacity & \citet{1984ApJ...285...89D} \\
            Mean dust grain size &  \num{3.7e-5}\ \text{cm} \\
            Dust-to-gas mass ratio & 0.01 \\
            Grazing angle & 0.05 \\
            Density power-law slope & $-1$ \\
            \noalign{\smallskip}
            \hline
        \end{tabular*}

   \end{table}

\begin{table}
\caption{Initial chemical abundances\label{tab:initial_abundances}}
\begin{tabular*}{\linewidth}{@{\extracolsep{\fill}}lc|lc}
\hline\hline
            \noalign{\smallskip}
        Species & Abundance &Species & Abundance \\
    & $n$(X) / $n$(H) & & $n$(X) / $n$(H) \\
            \noalign{\smallskip}
\hline
            \noalign{\smallskip}
ortho-H$_2$ & 0.375 &      CH$_4$ ice & \num{1.8e-5}\\
para-H$_2$ & 0.125  &       CO ice & \num{6.0e-5}\\
HD & \num{1.55e-5}  &       CH$_3$OH ice & \num{4.5e-5}\\
H & \num{5.0e-7}      &         CO$_2$ ice & \num{6.0e-5}\\
He & 0.098          &         CH$_4$ ice & \num{1.8e-5}\\
Cl & \num{1.0e-9}     &          H$_2$O ice & \num{3.0e-4}\\
Si & \num{8.0e-9}     &          H$_2$S ice & \num{6.6e-6}\\
Fe & \num{3.0e-9}     &          N$_2$ ice & \num{2.1e-5}\\
Mg & \num{7.0e-9}     &          NH$_3$ ice & \num{2.1e-5}\\
Na & \num{2.0e-9}     &
P & \num{2.0e-10}     \\
            \noalign{\smallskip}
\hline
\end{tabular*}
\end{table}

The disk physical structure in our models is defined through stellar mass, $M_{\rm star}$, disk mass, $M_{\rm disk}$, and the characteristic radius, $R_{\rm c}$; these parameters define the distribution of density, temperature, and radiation field in the (R, z) plane. The stellar mass governs the stellar temperature and luminosity, which are calculated at the age of 1\,Myr using the evolutionary  models by \citet{2008ASPC..387..189Y}. The X-ray radiation field is calculated using \citet{Bruderer_ea_2009a}. The interstellar cosmic ray ionization rate was calculated according to \citet{2018A&A...614A.111P}. We create an ensemble of 540 models with different stellar mass $M_{\rm star}$, disk mass, $M_{\rm disk}$, critical radius, $R_{\rm c}$, and stellar X-ray luminosity, $L_X$,  to cover a wide range of physical conditions typical for protoplanetary disks: $M_{\rm star} = 0.2 ...  1.0\,M_{\sun}$, $M_{\rm disk} = 0.1 ...  10\,\%\,M_{\rm star}$, $R_{c} = 20 ... 170$\,au, and $L_X = 10^{29 ... 31}\ $erg s$^{-1}$.
Each model includes 50 logarithmically spaced radial points in the range  of $R=0.1...1000$\,au and 80 vertical points in the range of $z/R = 0 ... 1$. The dust size distribution is described by a power law with a $-3.5$ exponent between \num{5e-7} and \num{2.5e-3}\,cm. The UV radiation field for photoreactions is calculated using dust opacities based on \citet{1984ApJ...285...89D}. An averaged grain size of \num{3.7e-5}\,cm is adopted for surface reaction rates.
The UV excess from accretion is defined as $L_{\rm acc} = 1.5\ G \dot{M}_{\rm acc} M_{\rm star} / R_{\rm star}$. The effective temperature of the accretion region is assumed to be 10000\,K. A summary of these parameters is presented in Table~\ref{tab:andesinp}. With 4000 spatial points in each model, we have $2\,160\,000$ points in total, sampling the chemical model output in conditions typical of protoplanetary disks. The total computing resource usage for the data generation took about 1 core year.

\begin{figure}
    \centering
    \includegraphics[width=\linewidth]{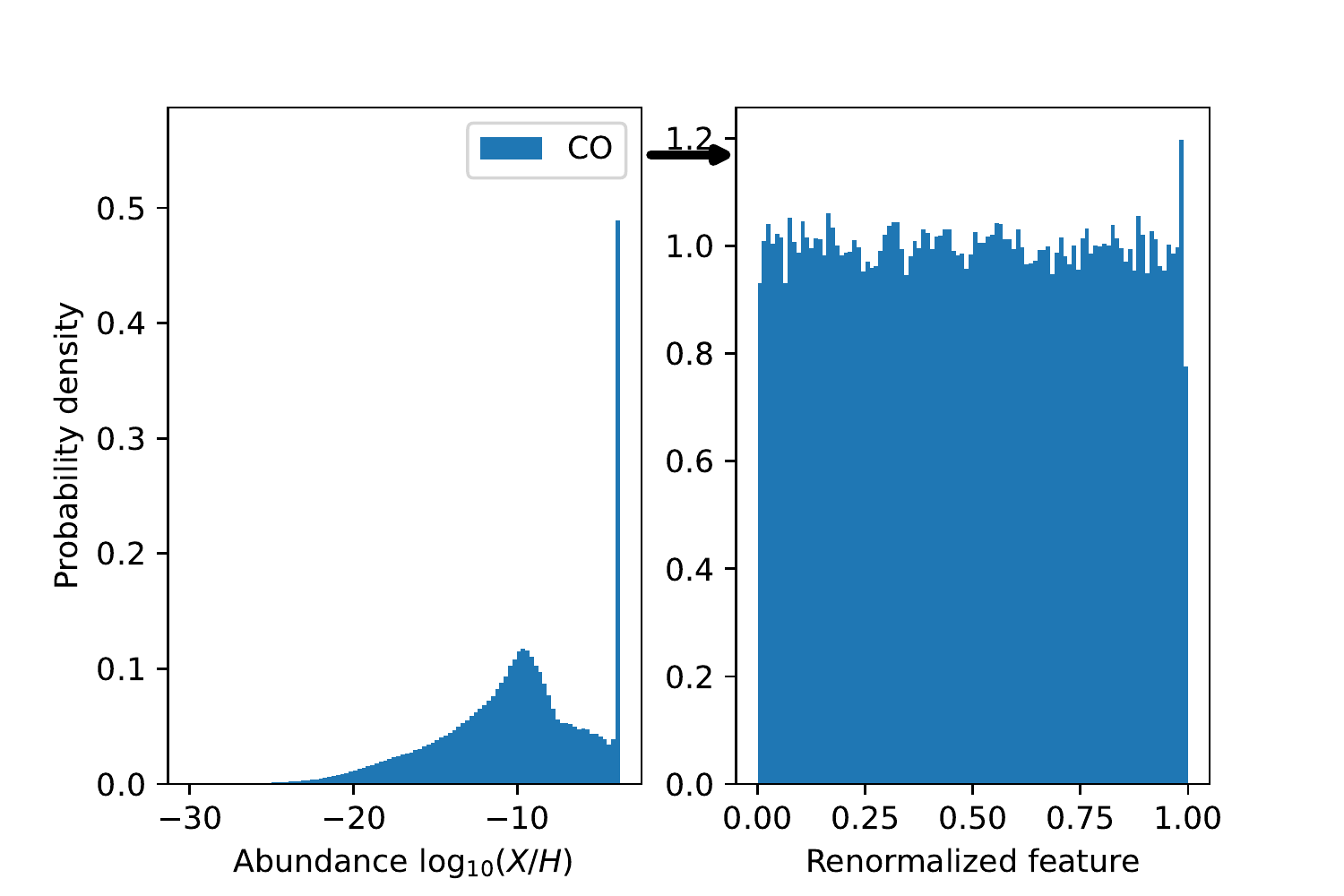}
    \caption{Example of the re-normalization of $\log_{10}X_{\rm CO}$ abundance prior to interpolation. The peak at $-4$ in the unnormalized data corresponds to the condition when all the available C and O form CO.}
    \label{fig:feature_normalization}
\end{figure}

\section{KNN estimator training}
\begin{figure*}[h]
    \centering
     ($\log_{10} X_c$) \hspace{3.0cm} (bias, dex) \hspace{2.5cm} (std, dex) \hspace{3.0cm} (histogram)\\
    \makebox[\textwidth][c]{\includegraphics[width=1.2\linewidth]{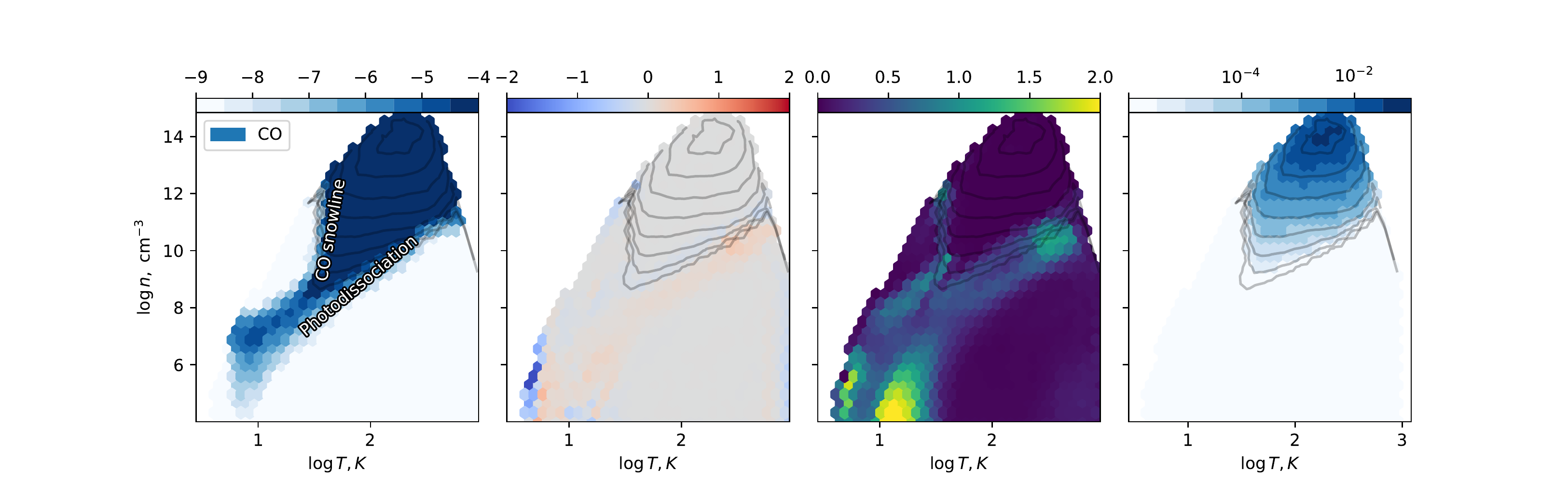}}
    (a) \hspace{3.9cm} (b) \hspace{3.9cm} (c) \hspace{3.9cm} (d)\\
    \caption{Performance of ML-accelerated chemistry predictions for CO. (a): mean $\log_{10}$ predicted relative abundance as a function of local temperature, gas density, and ionization rate. Darker areas correspond to larger relative (to H atoms) abundance. (b): median of the difference between the predicted values and test set data (bias, dex), in dex, as a function of temperature and density. Gray areas correspond to an unbiased fit. (c): the standard deviation between the predicted values and test set data, in dex (std, dex). (d): relative density (histogram) of species within the data points, with contours, which are also present on other panels. Various regions of the protoplanetary disk are described on panel (a). A detailed description of the processes leading to this figure is in the main text. Other molecules are shown in Fig.~\ref{fig:ions}.}
    \label{fig:performance}
\end{figure*}

First, we built a data frame using \texttt{pandas} \citep{pandas}, which contains a subset of local physical parameters and a set of selected chemical species. For demonstration purposes, we chose CO, \hcop{}, \dcop{}, and electrons, relevant to ionization studies \citep{2020A&A...644A...4S, 2021ApJS..257...13A}. We also demonstrate the possible applications to other species in Appendix~\ref{app:morespecies}. We have selected observationally-relevant disk positions with gas number density above $10^4$ cm$^{-3}$, resulting in $1\,183\,212$ data points. The subsequent analysis was performed using the \texttt{Scikit-learn} python library \citep{scikit-learn}. The physical quantities (input features) and the chemical abundances normalized to the total number of H atoms (output features) were renormalized to a uniform distribution in order to make the parameter space more uniformly sampled (see the example in Fig.~\ref{fig:feature_normalization}). We split the data into a training set (432 disks, 946586 points) and a test set (108 disks, 236626 points).  All the points from a single disk model must appear either in the training or a test set to avoid overfitting. The algorithm should predict the new points based on points from the other disks rather than interpolating nearby points of the same disk.

We used the \texttt{KNeighborsRegressor} estimator \citep{NearestNeighbors}. The algorithm finds the $k$ nearest data points (in the input feature space) and interpolates the output feature values between them. For robustness, we choose the median value of these data points. The value of $k$ should be chosen based on the data. If $k=1$, it is a classical "nearest" interpolation. This sort of interpolation is very sensitive to outliers (overfitting) which, in our case, can represent a rare combination of parameters or even a numerical failure of the original chemical kinetics solver within ANDES. If $k$ is too large, then the local behavior of chemistry cannot be properly caught.  In the worst case, where $k$ equals the number of data points, the solution would simply represent their median value.

We used cross-validation to choose the value of $k$ and ensure the quality of the interpolation. In this approach, the training set is divided into 10 parts (splits), with all points from each disk being in the same split. The estimator is trained on 9/10 of the training data set and its performance is benchmarked against the remaining part, using the square sum of errors metric in the renormalized space to quantify the quality of the fit. This is repeated for each split, and the average performance is estimated. Then the same procedure repeats for another value of $k$, and this way, the best-performing value of $k$ is found. The estimator with the best $k$ is fitted again afterward on the entire training set. The results can be saved as a python binary ("pickle") file and used as a fast callable function in other applications.

\section{Results and discussion}

First, we chose the local gas density, dust temperature, and ionization rate (the minimum set of key physical parameters for chemistry) as input features. We provide an example of the performance of the fit in Fig.~\ref{fig:performance}, using the gas-phase CO molecule. As seen in panel (a), CO is abundant in dense regions with a temperature above 30~K. Typically, the CO snowline should be at around 20~K. However, in our modeling, CO is also absent in the gas at higher temperatures due to the chemical transformation into CO$_2$ on dust grain surfaces \citep{2017ApJ...849..130M,2018A&A...618A.182B}. At lower densities, which correspond to the outer disk, photodesorption by the interstellar UV and ionizing radiation maintain some amount of CO in the gas phase, enough for self-shielding. Low-density and high-temperature areas belong to the disk atmosphere, where the UV-radiation destroys CO. The fit reproduces this general behavior, showing insignificant systematic error (bias, dex) on the panel (b). The fit scatter (panel c) is especially low ($<0.5$ dex) for the inner disk and midplane. Most of the disk CO gas is present in the inner disk and, hence, the fit reproduces the majority of the gas-phase CO in disks with a reasonable level of accuracy. Moreover, the fit correctly predicts low CO abundances outside the CO snowline and in the disk atmosphere. Significant scatter (above 1 dex) is present only at the radiation-sensitive transition zone between the atmosphere and the rest of the disk in the low-density area. The top-left corner (very high density, very low temperature) is not covered by the original data set, as regions with such conditions never appeared in the disk model grid.

 Notably, we can reproduce deuterium fractionation. In  Fig.~\ref{fig:ions}, we show other species included in the estimator's output: \hcop{}, \dcop{}, and \electr{}. While electron density is fitted almost perfectly, the \hcop{} and \dcop{} fits show $\sim$1~dex scatter in the transition zone between the inner disk and the photodissociation area, where just a small amount of gas is present. Overall, with just three inputs' feature set,  the estimator is able to predict the disk chemistry with a good accuracy below 0.5 dex for most of the parameter space.

The addition of more input parameters to the input features further increases the quality of the fit. In Fig.~\ref{fig:ions4}, we show the model after adding local UV radiation intensity to the set of the input features. We can see a significantly lower scatter in the whole parameter space. Temperature and density remain the best predictors of the disk chemistry, explaining the major variations of $\gg 5$ dex, with the local ionization rate contributing to $\sim1-2$ dex and ultraviolet field contributing to $\sim1$ dex in relevant disk regions. The impact of ionizing radiation is more important than the UV for the molecules, as they reside in the deeper layers of the disk, while X-ray and cosmic rays penetrate deeper towards the midplane. Even in the three-parameter fit, it is important to take the local UV radiation field into account for the data generation process. Nevertheless, as the UV field is correlated with the combination of other input parameters, it is not necessary to have it as an input feature for the KNN algorithm.

\section{Conclusions}
   We applied, for the first time, a machine-learning estimator to physical-chemical protoplanetary disk models to predict chemical abundances much more quickly than traditional "full" chemical calculations. Our estimator uses a small and easily-calculated set of local physical parameters as input features: the dust temperature, density, and ionization rate, with the possible addition of a local UV radiation strength. We applied this method to a pre-computed database of 540 protoplanetary disks of various masses, sizes, and stellar properties, including X-ray luminosities. We studied the effectiveness and limitations of the method due to the small input feature set, demonstrating how the addition of a local UV field improves the accuracy for four species in the gas: CO, \hcop{}, \dcop{}, and \electr{}.

    This approach is general and can be applied not only to this set of species and ANDES thermo-chemical disk models, but also to other species and astrochemical applications. For such purposes, the outputs of other astrochemical models \citep[e.g.,][]{Bruderer_ea_2009a,Woitke_Kamp_Thi_2009_ProDiMo} with relevant physical parameters and desired chemical species abundances can be processed with the same approach. We publish the ANDES-generated data and the Jupyter notebook to reproduce our results on GitHub \footnote{https://github.com/SmirnGreg/diskchef\_chemistry}. 
    These results will be used in an upcoming series of papers to fit the molecular data obtained in the framework of the large observing program on NOEMA, L19ME (PI: Th. Henning). We use this method to rapidly calculate chemical composition for the Bayesian retrieval of disk physical parameters using CO isotopologues (Francheschi et al., in prep.) and a combined fit of CO and \hcop{} isotopologues (Smirnov-Pinchukov et al., in prep).

\begin{acknowledgements}
    The authors acknowledge the contribution of the Python open-source community for providing high-quality data analysis tools. All figures were created using \texttt{matplotlib} \citep{Hunter:2007}. GSP thanks Morgan Fouesneau, Ivelina Momcheva, and Markus Schmalzl for discussions about machine learning at MPIA. TM and VA were supported by the grant 075-15-2020-780 (N13.1902.21.0039) of Ministry of Science and Higher Education of the Russian Federation. TH and DS acknowledge support from the European Research Council under the Horizon 2020 Framework Program via the ERC Advanced Grant Origins 83 24 28.
\end{acknowledgements}

%
  \bibliographystyle{aa} 
  \bibliography{paper} 

\begin{thebibliography}{50}
\expandafter\ifx\csname natexlab\endcsname\relax\def\natexlab#1{#1}\fi

\bibitem[{{Aikawa} {et~al.}(2021){Aikawa}, {Cataldi}, {Yamato}, {Zhang},
  {Booth}, {Furuya}, {Andrews}, {Bae}, {Bergin}, {Bergner}, {Bosman},
  {Cleeves}, {Czekala}, {Guzm{\'a}n}, {Huang}, {Ilee}, {Law}, {Le Gal},
  {Loomis}, {M{\'e}nard}, {Nomura}, {{\"O}berg}, {Qi}, {Schwarz}, {Teague},
  {Tsukagoshi}, {Walsh}, \& {Wilner}}]{2021ApJS..257...13A}
{Aikawa}, Y., {Cataldi}, G., {Yamato}, Y., {et~al.} 2021, \apjs, 257, 13

\bibitem[{{Aikawa} {et~al.}(2002){Aikawa}, {van Zadelhoff}, {van Dishoeck}, \&
  {Herbst}}]{2002A&A...386..622A}
{Aikawa}, Y., {van Zadelhoff}, G.~J., {van Dishoeck}, E.~F., \& {Herbst}, E.
  2002, \aap, 386, 622

\bibitem[{{Akimkin} {et~al.}(2013){Akimkin}, {Zhukovska}, {Wiebe}, {Semenov},
  {Pavlyuchenkov}, {Vasyunin}, {Birnstiel}, \& {Henning}}]{2013ApJ...766....8A}
{Akimkin}, V., {Zhukovska}, S., {Wiebe}, D., {et~al.} 2013, \apj, 766, 8

\bibitem[{{Albertsson} {et~al.}(2014){Albertsson}, {Semenov}, \&
  {Henning}}]{2014ApJ...784...39A}
{Albertsson}, T., {Semenov}, D., \& {Henning}, T. 2014, \apj, 784, 39

\bibitem[{{Albertsson} {et~al.}(2013){Albertsson}, {Semenov}, {Vasyunin},
  {Henning}, \& {Herbst}}]{2013ApJS..207...27A}
{Albertsson}, T., {Semenov}, D.~A., {Vasyunin}, A.~I., {Henning}, T., \&
  {Herbst}, E. 2013, \apjs, 207, 27

\bibitem[{{Ansdell} {et~al.}(2016){Ansdell}, {Williams}, {van der Marel},
  {Carpenter}, {Guidi}, {Hogerheijde}, {Mathews}, {Manara}, {Miotello},
  {Natta}, {Oliveira}, {Tazzari}, {Testi}, {van Dishoeck}, \& {van
  Terwisga}}]{2016ApJ...828...46A}
{Ansdell}, M., {Williams}, J.~P., {van der Marel}, N., {et~al.} 2016, \apj,
  828, 46

\bibitem[{{Ard{\'e}vol Mart{\'\i}nez} {et~al.}(2022){Ard{\'e}vol
  Mart{\'\i}nez}, {Min}, {Kamp}, \& {Palmer}}]{2022A&A...662A.108A}
{Ard{\'e}vol Mart{\'\i}nez}, F., {Min}, M., {Kamp}, I., \& {Palmer}, P.~I.
  2022, \aap, 662, A108

\bibitem[{{Bosman} {et~al.}(2018){Bosman}, {Walsh}, \& {van
  Dishoeck}}]{2018A&A...618A.182B}
{Bosman}, A.~D., {Walsh}, C., \& {van Dishoeck}, E.~F. 2018, \aap, 618, A182

\bibitem[{{Bruderer} {et~al.}(2009){Bruderer}, {Doty}, \&
  {Benz}}]{Bruderer_ea_2009a}
{Bruderer}, S., {Doty}, S.~D., \& {Benz}, A.~O. 2009, \apjs, 183, 179

\bibitem[{{Carleo} {et~al.}(2019){Carleo}, {Cirac}, {Cranmer}, {Daudet},
  {Schuld}, {Tishby}, {Vogt-Maranto}, \& {Zdeborov{\'a}}}]{2019RvMP...91d5002C}
{Carleo}, G., {Cirac}, I., {Cranmer}, K., {et~al.} 2019, Reviews of Modern
  Physics, 91, 045002

\bibitem[{{Cieza} {et~al.}(2019){Cieza}, {Ru{\'\i}z-Rodr{\'\i}guez}, {Hales},
  {Casassus}, {P{\'e}rez}, {Gonzalez-Ruilova}, {C{\'a}novas}, {Williams},
  {Zurlo}, {Ansdell}, {Avenhaus}, {Bayo}, {Bertrang}, {Christiaens}, {Dent},
  {Ferrero}, {Gamen}, {Olofsson}, {Orcajo}, {Pe{\~n}a Ram{\'\i}rez},
  {Principe}, {Schreiber}, \& {van der Plas}}]{2019MNRAS.482..698C}
{Cieza}, L.~A., {Ru{\'\i}z-Rodr{\'\i}guez}, D., {Hales}, A., {et~al.} 2019,
  \mnras, 482, 698

\bibitem[{{Dieleman} {et~al.}(2015){Dieleman}, {Willett}, \&
  {Dambre}}]{2015MNRAS.450.1441D}
{Dieleman}, S., {Willett}, K.~W., \& {Dambre}, J. 2015, \mnras, 450, 1441

\bibitem[{{Draine} \& {Lee}(1984)}]{1984ApJ...285...89D}
{Draine}, B.~T. \& {Lee}, H.~M. 1984, \apj, 285, 89

\bibitem[{{Dunjko} \& {Briegel}(2018)}]{2018RPPh...81g4001D}
{Dunjko}, V. \& {Briegel}, H.~J. 2018, Reports on Progress in Physics, 81,
  074001

\bibitem[{{Eistrup} {et~al.}(2016){Eistrup}, {Walsh}, \& {van
  Dishoeck}}]{2016A&A...595A..83E}
{Eistrup}, C., {Walsh}, C., \& {van Dishoeck}, E.~F. 2016, \aap, 595, A83

\bibitem[{{Fedele} \& {Favre}(2020)}]{2020A&A...638A.110F}
{Fedele}, D. \& {Favre}, C. 2020, \aap, 638, A110

\bibitem[{{Garrod}(2013)}]{2013ApJ...765...60G}
{Garrod}, R.~T. 2013, \apj, 765, 60

\bibitem[{{Garufi} {et~al.}(2020){Garufi}, {Podio}, {Codella}, {Rygl},
  {Bacciotti}, {Facchini}, {Fedele}, {Miotello}, {Teague}, \&
  {Testi}}]{2020A&A...636A..65G}
{Garufi}, A., {Podio}, L., {Codella}, C., {et~al.} 2020, \aap, 636, A65

\bibitem[{Goldberger {et~al.}(2005)Goldberger, Hinton, Roweis, \&
  Salakhutdinov}]{NearestNeighbors}
Goldberger, J., Hinton, G.~E., Roweis, S., \& Salakhutdinov, R.~R. 2005, in
  Advances in Neural Information Processing Systems, ed. L.~Saul, Y.~Weiss, \&
  L.~Bottou, Vol.~17 (MIT Press)

\bibitem[{{Grassi} {et~al.}(2014){Grassi}, {Bovino}, {Schleicher}, {Prieto},
  {Seifried}, {Simoncini}, \& {Gianturco}}]{2014MNRAS.439.2386G}
{Grassi}, T., {Bovino}, S., {Schleicher}, D.~R.~G., {et~al.} 2014, \mnras, 439,
  2386

\bibitem[{{Grassi} {et~al.}(2021){Grassi}, {Nauman}, {Ramsey}, {Bovino},
  {Picogna}, \& {Ercolano}}]{2021arXiv210409516G}
{Grassi}, T., {Nauman}, F., {Ramsey}, J.~P., {et~al.} 2021, arXiv e-prints,
  arXiv:2104.09516

\bibitem[{{Guzm{\'a}n} {et~al.}(2021){Guzm{\'a}n}, {Bergner}, {Law},
  {{\"O}berg}, {Walsh}, {Cataldi}, {Aikawa}, {Bergin}, {Czekala}, {Huang},
  {Andrews}, {Loomis}, {Zhang}, {Le Gal}, {Alarc{\'o}n}, {Ilee}, {Teague},
  {Cleeves}, {Wilner}, {Long}, {Schwarz}, {Bosman}, {P{\'e}rez}, {M{\'e}nard},
  \& {Liu}}]{2021ApJS..257....6G}
{Guzm{\'a}n}, V.~V., {Bergner}, J.~B., {Law}, C.~J., {et~al.} 2021, \apjs, 257,
  6

\bibitem[{{Holdship} \& {Viti}(2022)}]{2022A&A...658A.103H}
{Holdship}, J. \& {Viti}, S. 2022, \aap, 658, A103

\bibitem[{{Holdship} {et~al.}(2021){Holdship}, {Viti}, {Haworth}, \&
  {Ilee}}]{2021A&A...653A..76H}
{Holdship}, J., {Viti}, S., {Haworth}, T.~J., \& {Ilee}, J.~D. 2021, \aap, 653,
  A76

\bibitem[{{Holdship} {et~al.}(2017){Holdship}, {Viti}, {Jim{\'e}nez-Serra},
  {Makrymallis}, \& {Priestley}}]{2017AJ....154...38H}
{Holdship}, J., {Viti}, S., {Jim{\'e}nez-Serra}, I., {Makrymallis}, A., \&
  {Priestley}, F. 2017, \aj, 154, 38

\bibitem[{Hunter(2007)}]{Hunter:2007}
Hunter, J.~D. 2007, Computing in Science \& Engineering, 9, 90

\bibitem[{{Ilee} {et~al.}(2021){Ilee}, {Walsh}, {Booth}, {Aikawa}, {Andrews},
  {Bae}, {Bergin}, {Bergner}, {Bosman}, {Cataldi}, {Cleeves}, {Czekala},
  {Guzm{\'a}n}, {Huang}, {Law}, {Le Gal}, {Loomis}, {M{\'e}nard}, {Nomura},
  {{\"O}berg}, {Qi}, {Schwarz}, {Teague}, {Tsukagoshi}, {Wilner}, {Yamato}, \&
  {Zhang}}]{2021ApJS..257....9I}
{Ilee}, J.~D., {Walsh}, C., {Booth}, A.~S., {et~al.} 2021, \apjs, 257, 9

\bibitem[{{Keil} {et~al.}(2022){Keil}, {Viti}, \&
  {Holdship}}]{2022ApJ...927..203K}
{Keil}, M., {Viti}, S., \& {Holdship}, J. 2022, \apj, 927, 203

\bibitem[{{Lee} {et~al.}(2021){Lee}, {Patterson}, {Burkhardt}, {Vankayalapati},
  {McCarthy}, \& {McGuire}}]{2021ApJ...917L...6L}
{Lee}, K. L.~K., {Patterson}, J., {Burkhardt}, A.~M., {et~al.} 2021, \apjl,
  917, L6

\bibitem[{{Molyarova} {et~al.}(2018){Molyarova}, {Akimkin}, {Semenov},
  {{\'A}brah{\'a}m}, {Henning}, {K{\'o}sp{\'a}l}, {Vorobyov}, \&
  {Wiebe}}]{2018ApJ...866...46M}
{Molyarova}, T., {Akimkin}, V., {Semenov}, D., {et~al.} 2018, \apj, 866, 46

\bibitem[{{Molyarova} {et~al.}(2017){Molyarova}, {Akimkin}, {Semenov},
  {Henning}, {Vasyunin}, \& {Wiebe}}]{2017ApJ...849..130M}
{Molyarova}, T., {Akimkin}, V., {Semenov}, D., {et~al.} 2017, \apj, 849, 130

\bibitem[{{{\"O}berg} {et~al.}(2011){{\"O}berg}, {Boogert}, {Pontoppidan}, {van
  den Broek}, {van Dishoeck}, {Bottinelli}, {Blake}, \&
  {Evans}}]{2011ApJ...740..109O}
{{\"O}berg}, K.~I., {Boogert}, A.~C.~A., {Pontoppidan}, K.~M., {et~al.} 2011,
  \apj, 740, 109

\bibitem[{{{\"O}berg} {et~al.}(2021){{\"O}berg}, {Guzm{\'a}n}, {Walsh},
  {Aikawa}, {Bergin}, {Law}, {Loomis}, {Alarc{\'o}n}, {Andrews}, {Bae},
  {Bergner}, {Boehler}, {Booth}, {Bosman}, {Calahan}, {Cataldi}, {Cleeves},
  {Czekala}, {Furuya}, {Huang}, {Ilee}, {Kurtovic}, {Le Gal}, {Liu}, {Long},
  {M{\'e}nard}, {Nomura}, {P{\'e}rez}, {Qi}, {Schwarz}, {Sierra}, {Teague},
  {Tsukagoshi}, {Yamato}, {van't Hoff}, {Waggoner}, {Wilner}, \&
  {Zhang}}]{2021ApJS..257....1O_MAPS1}
{{\"O}berg}, K.~I., {Guzm{\'a}n}, V.~V., {Walsh}, C., {et~al.} 2021, \apjs,
  257, 1

\bibitem[{{Padovani} {et~al.}(2018){Padovani}, {Ivlev}, {Galli}, \&
  {Caselli}}]{2018A&A...614A.111P}
{Padovani}, M., {Ivlev}, A.~V., {Galli}, D., \& {Caselli}, P. 2018, \aap, 614,
  A111

\bibitem[{Pedregosa {et~al.}(2011)Pedregosa, Varoquaux, Gramfort, Michel,
  Thirion, Grisel, Blondel, Prettenhofer, Weiss, Dubourg, Vanderplas, Passos,
  Cournapeau, Brucher, Perrot, \& Duchesnay}]{scikit-learn}
Pedregosa, F., Varoquaux, G., Gramfort, A., {et~al.} 2011, Journal of Machine
  Learning Research, 12, 2825

\bibitem[{{Pegues} {et~al.}(2021){Pegues}, {{\"O}berg}, {Bergner}, {Huang},
  {Pascucci}, {Teague}, {Andrews}, {Bergin}, {Cleeves}, {Guzm{\'a}n}, {Long},
  {Qi}, \& {Wilner}}]{2021ApJ...911..150P}
{Pegues}, J., {{\"O}berg}, K.~I., {Bergner}, J.~B., {et~al.} 2021, \apj, 911,
  150

\bibitem[{{Pegues} {et~al.}(2020){Pegues}, {{\"O}berg}, {Bergner}, {Loomis},
  {Qi}, {Le Gal}, {Cleeves}, {Guzm{\'a}n}, {Huang}, {J{\o}rgensen}, {Andrews},
  {Blake}, {Carpenter}, {Schwarz}, {Williams}, \&
  {Wilner}}]{2020ApJ...890..142P}
{Pegues}, J., {{\"O}berg}, K.~I., {Bergner}, J.~B., {et~al.} 2020, \apj, 890,
  142

\bibitem[{Reback {et~al.}(2021)Reback, McKinney, jbrockmendel, den Bossche,
  Augspurger, Cloud, Hawkins, gfyoung, Sinhrks, Roeschke, Klein, Petersen,
  Tratner, She, Ayd, Naveh, patrick, Garcia, Schendel, Hayden, Saxton,
  Jancauskas, Gorelli, Shadrach, McMaster, Battiston, Seabold, Dong, chris b1,
  \& h~vetinari}]{pandas}
Reback, J., McKinney, W., jbrockmendel, {et~al.} 2021, pandas-dev/pandas:
  Pandas 1.2.4

\bibitem[{{Ribas} {et~al.}(2020){Ribas}, {Espaillat}, {Mac{\'\i}as}, \&
  {Sarro}}]{2020A&A...642A.171R}
{Ribas}, {\'A}., {Espaillat}, C.~C., {Mac{\'\i}as}, E., \& {Sarro}, L.~M. 2020,
  \aap, 642, A171

\bibitem[{{Ruaud} {et~al.}(2016){Ruaud}, {Wakelam}, \&
  {Hersant}}]{2016MNRAS.459.3756R}
{Ruaud}, M., {Wakelam}, V., \& {Hersant}, F. 2016, \mnras, 459, 3756

\bibitem[{{Semenov} {et~al.}(2010){Semenov}, {Hersant}, {Wakelam}, {Dutrey},
  {Chapillon}, {Guilloteau}, {Henning}, {Launhardt}, {Pi{\'e}tu}, \&
  {Schreyer}}]{2010A&A...522A..42S}
{Semenov}, D., {Hersant}, F., {Wakelam}, V., {et~al.} 2010, \aap, 522, A42

\bibitem[{{Semenov} \& {Wiebe}(2011)}]{2011ApJS..196...25S}
{Semenov}, D. \& {Wiebe}, D. 2011, \apjs, 196, 25

\bibitem[{{Smirnov-Pinchukov} {et~al.}(2020){Smirnov-Pinchukov}, {Semenov},
  {Akimkin}, \& {Henning}}]{2020A&A...644A...4S}
{Smirnov-Pinchukov}, G.~V., {Semenov}, D.~A., {Akimkin}, V.~V., \& {Henning},
  T. 2020, \aap, 644, A4

\bibitem[{{Vasyunin} {et~al.}(2009){Vasyunin}, {Semenov}, {Wiebe}, \&
  {Henning}}]{2009ApJ...691.1459V}
{Vasyunin}, A.~I., {Semenov}, D.~A., {Wiebe}, D.~S., \& {Henning}, T. 2009,
  \apj, 691, 1459

\bibitem[{{Villadsen} {et~al.}(2022){Villadsen}, {Ligterink}, \&
  {Andersen}}]{2022arXiv220703906V}
{Villadsen}, T., {Ligterink}, N. F.~W., \& {Andersen}, M. 2022, arXiv e-prints,
  arXiv:2207.03906

\bibitem[{{Willacy} {et~al.}(1998){Willacy}, {Klahr}, {Millar}, \&
  {Henning}}]{1998A&A...338..995W}
{Willacy}, K., {Klahr}, H.~H., {Millar}, T.~J., \& {Henning}, T. 1998, \aap,
  338, 995

\bibitem[{{Williams} \& {Best}(2014)}]{WilliamsBest2014}
{Williams}, J.~P. \& {Best}, W. M.~J. 2014, \apj, 788, 59

\bibitem[{{Woitke} {et~al.}(2009){Woitke}, {Kamp}, \&
  {Thi}}]{Woitke_Kamp_Thi_2009_ProDiMo}
{Woitke}, P., {Kamp}, I., \& {Thi}, W.~F. 2009, \aap, 501, 383

\bibitem[{{Yorke} \& {Bodenheimer}(2008)}]{2008ASPC..387..189Y}
{Yorke}, H.~W. \& {Bodenheimer}, P. 2008, in Astronomical Society of the
  Pacific Conference Series, Vol. 387, Massive Star Formation: Observations
  Confront Theory, ed. H.~{Beuther}, H.~{Linz}, \& T.~{Henning}, 189

\bibitem[{{Yu} {et~al.}(2016){Yu}, {Willacy}, {Dodson-Robinson}, {Turner}, \&
  {Evans}}]{2016ApJ...822...53Y}
{Yu}, M., {Willacy}, K., {Dodson-Robinson}, S.~E., {Turner}, N.~J., \& {Evans},
  Neal~J., I. 2016, \apj, 822, 53

\end{thebibliography}
%
\begin{appendix} 

\section{Additional figures} 
\begin{figure*}
    \centering
     ($\log_{10} X_c$) \hspace{2.0cm} (bias, dex) \hspace{2.0cm} (std, dex) \hspace{2.0cm} (histogram)\\

    \makebox[\textwidth][c]{\includegraphics[width=1.0\linewidth]{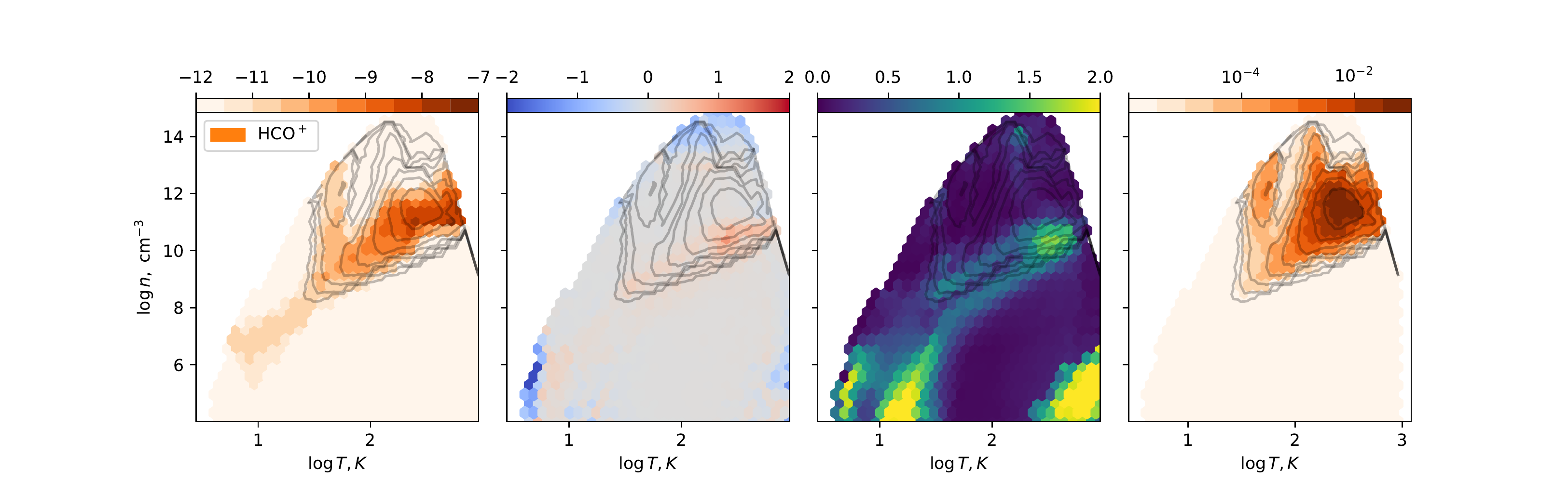}}
    \makebox[\textwidth][c]{\includegraphics[width=1.0\linewidth]{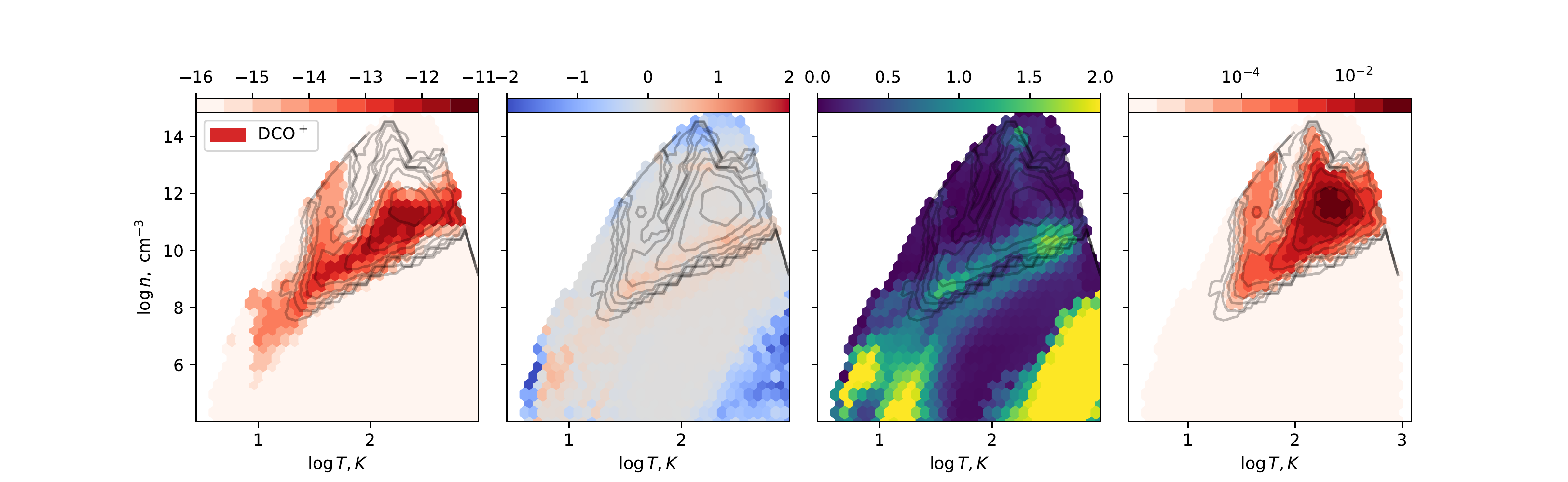}}
    \makebox[\textwidth][c]{\includegraphics[width=1.0\linewidth]{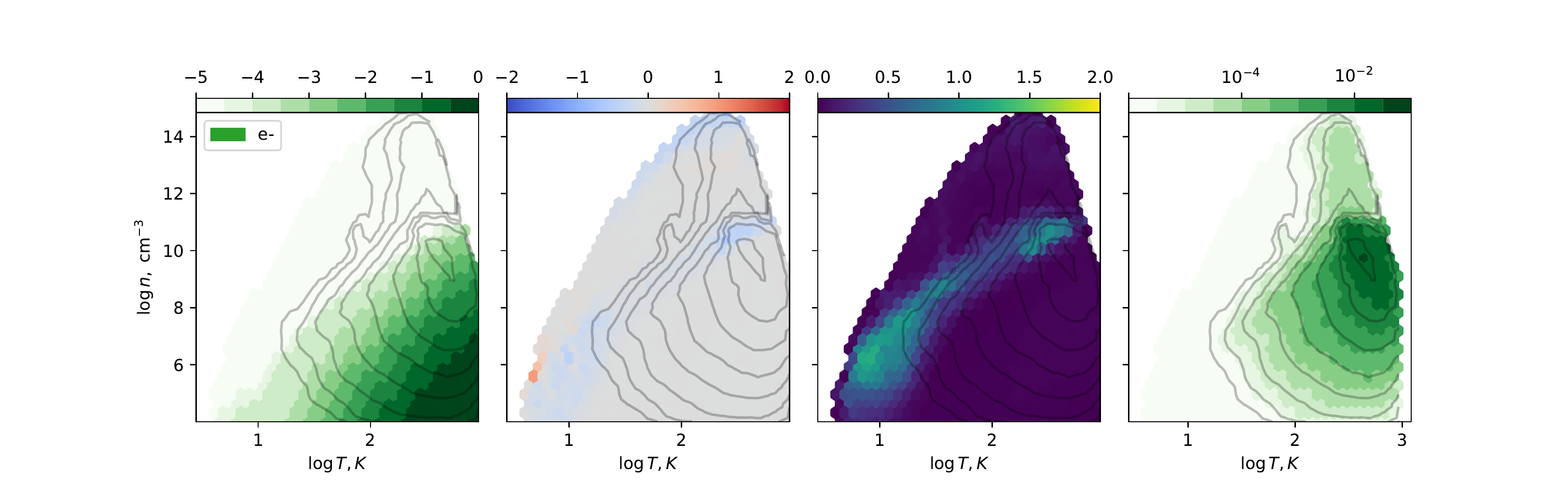}}
    \caption{Performance of ML-accelerated chemistry predictions for \hcop{}, \dcop{}, and electrons.}
    \label{fig:ions}
\end{figure*}

\begin{figure*}
    \centering
     ($\log_{10} X_c$) \hspace{2.0cm} (bias, dex) \hspace{2.0cm} (std, dex) \hspace{2.0cm} (histogram)\\
    \makebox[\textwidth][c]{\includegraphics[width=1.0\linewidth]{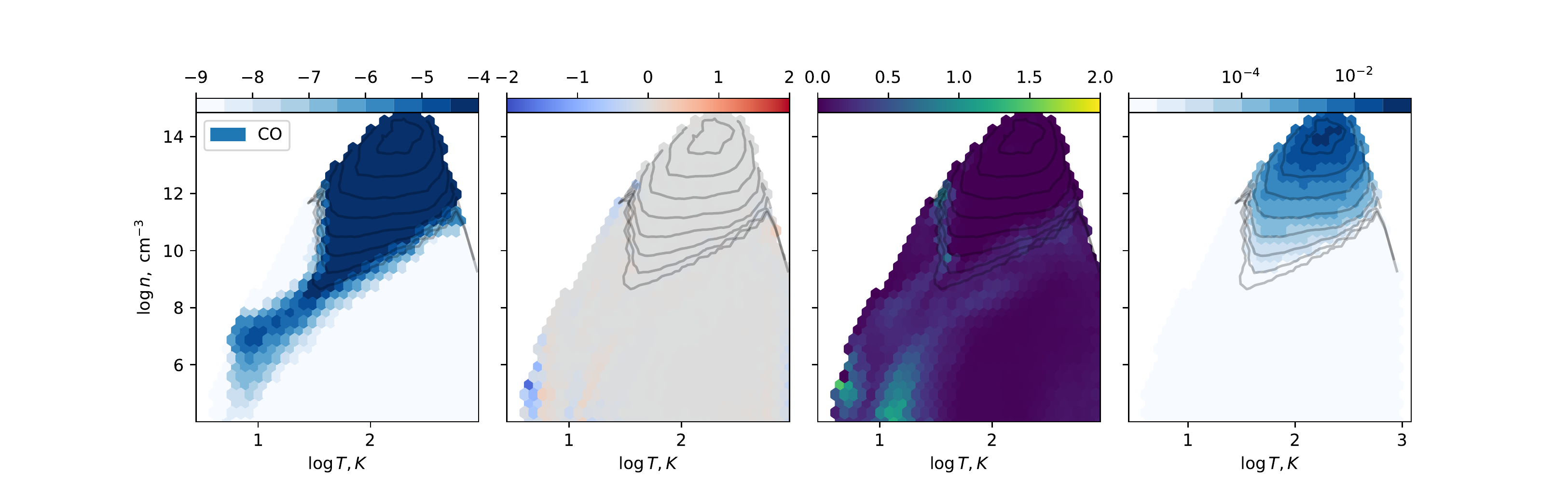}}
    \makebox[\textwidth][c]{\includegraphics[width=1.0\linewidth]{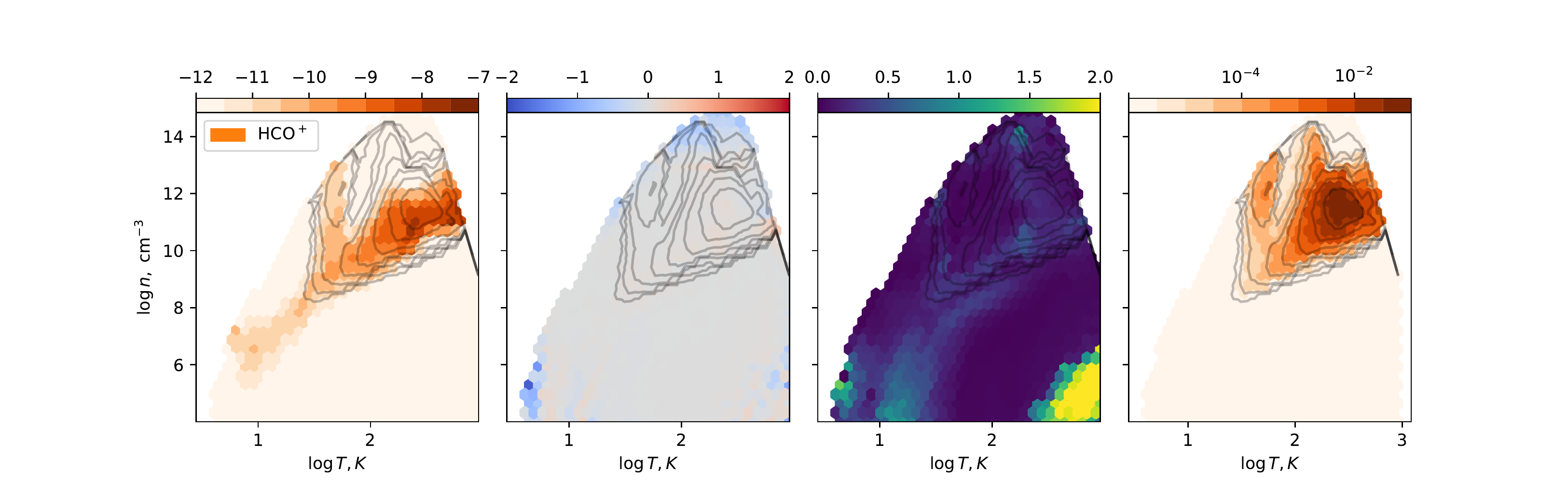}}
    \makebox[\textwidth][c]{\includegraphics[width=1.0\linewidth]{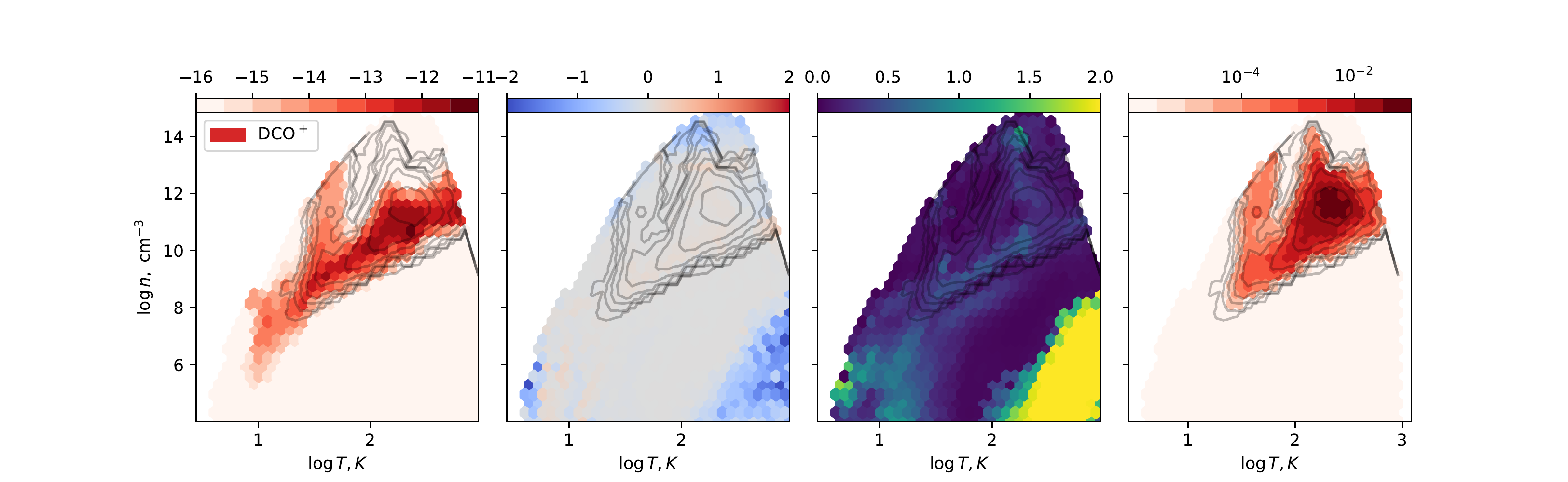}}
    \makebox[\textwidth][c]{\includegraphics[width=1.0\linewidth]{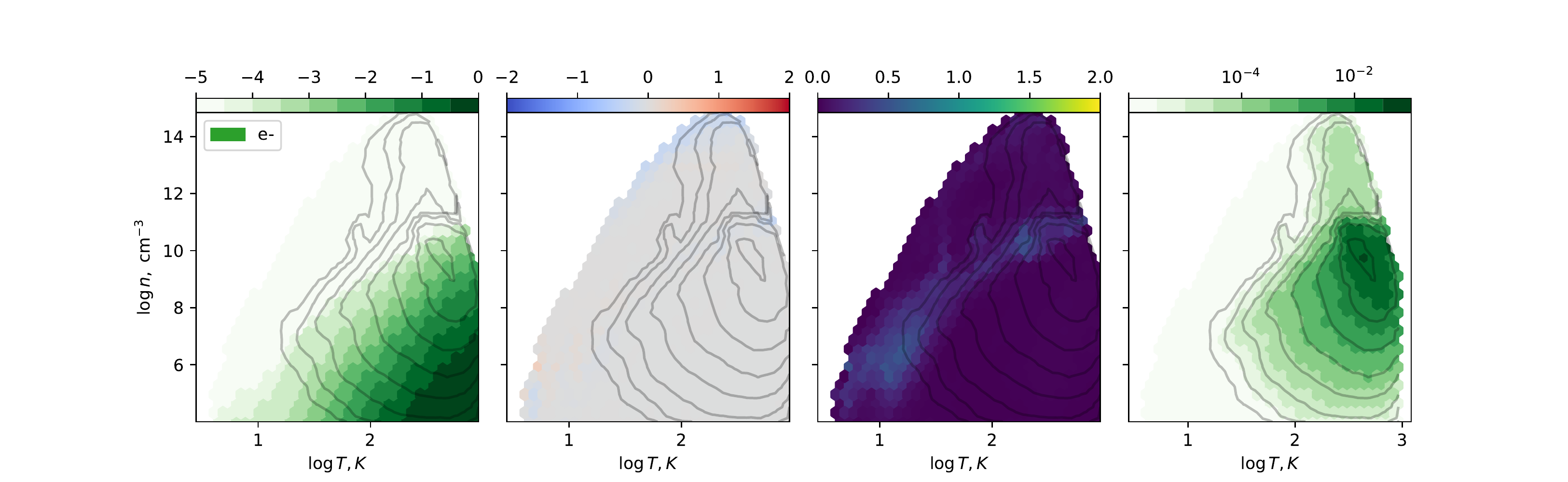}}
    \caption{Effect of adding UV radiation strength to the set of input features. The panels show the performance of ML-accelerated chemistry predictions, as in Fig.~\ref{fig:performance}. Adding UV radiation strength increases the accuracy of the fit, but only slightly in the areas of the parameter space dominated by the selected molecules. Depending on the molecular species and constraints on calculation time, it is not necessary to use UV as a parameter.}
    \label{fig:ions4}
\end{figure*}

\label{app:morespecies}
Fig.~\ref{fig:ions} demonstrates the application of the method on \hcop{}, \dcop{}, and electrons for three input features: local temperature, gas density, ionization rate. On the Fig.~\ref{fig:ions4} we present the result for the same molecules and CO with the local UV radiation strength added to the features list. In addition, we used our study to demonstrate the summary of the same method application to a larger number of different species, with four input features in Fig.~\ref{fig:morespecies}.

\begin{figure*}
\centering
     ($\log_{10} X_c$) \hspace{1.6cm} (std, dex) \hspace{2cm} ($\log_{10} X_c$) \hspace{1.6cm} (std, dex)\\
    \includegraphics[width=0.37\linewidth]{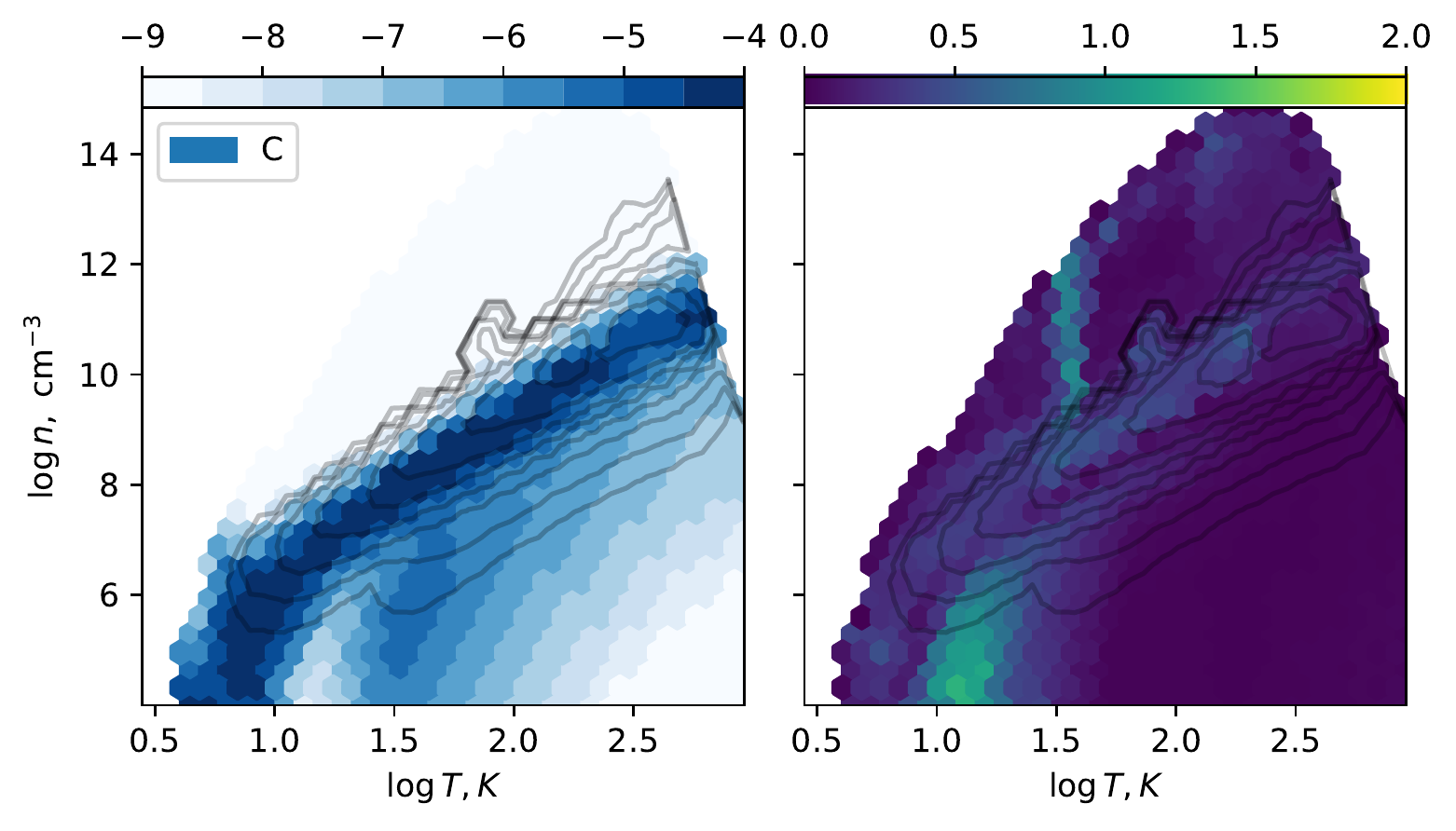}
    \includegraphics[width=0.37\linewidth]{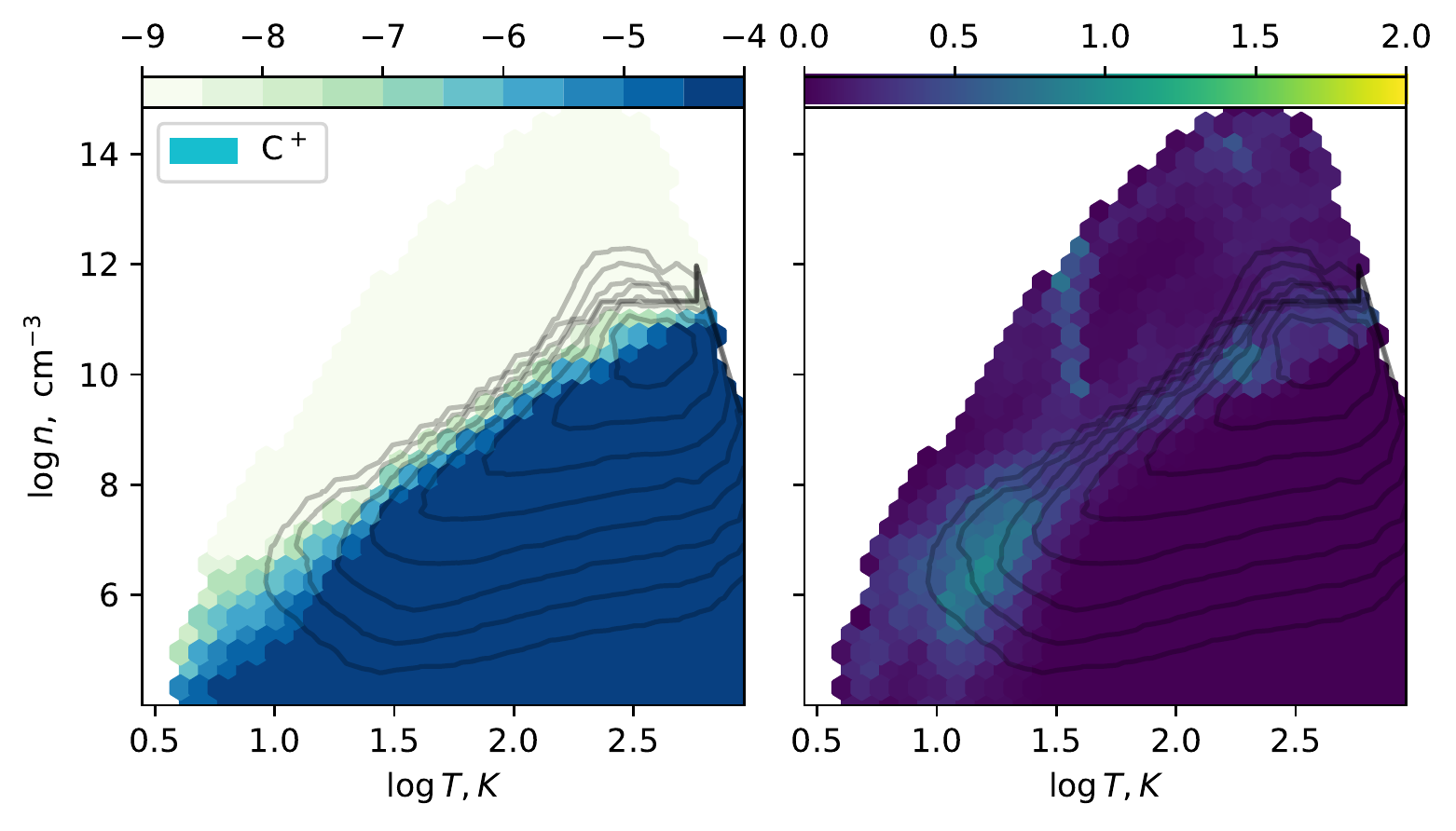}

    \includegraphics[width=0.37\linewidth]{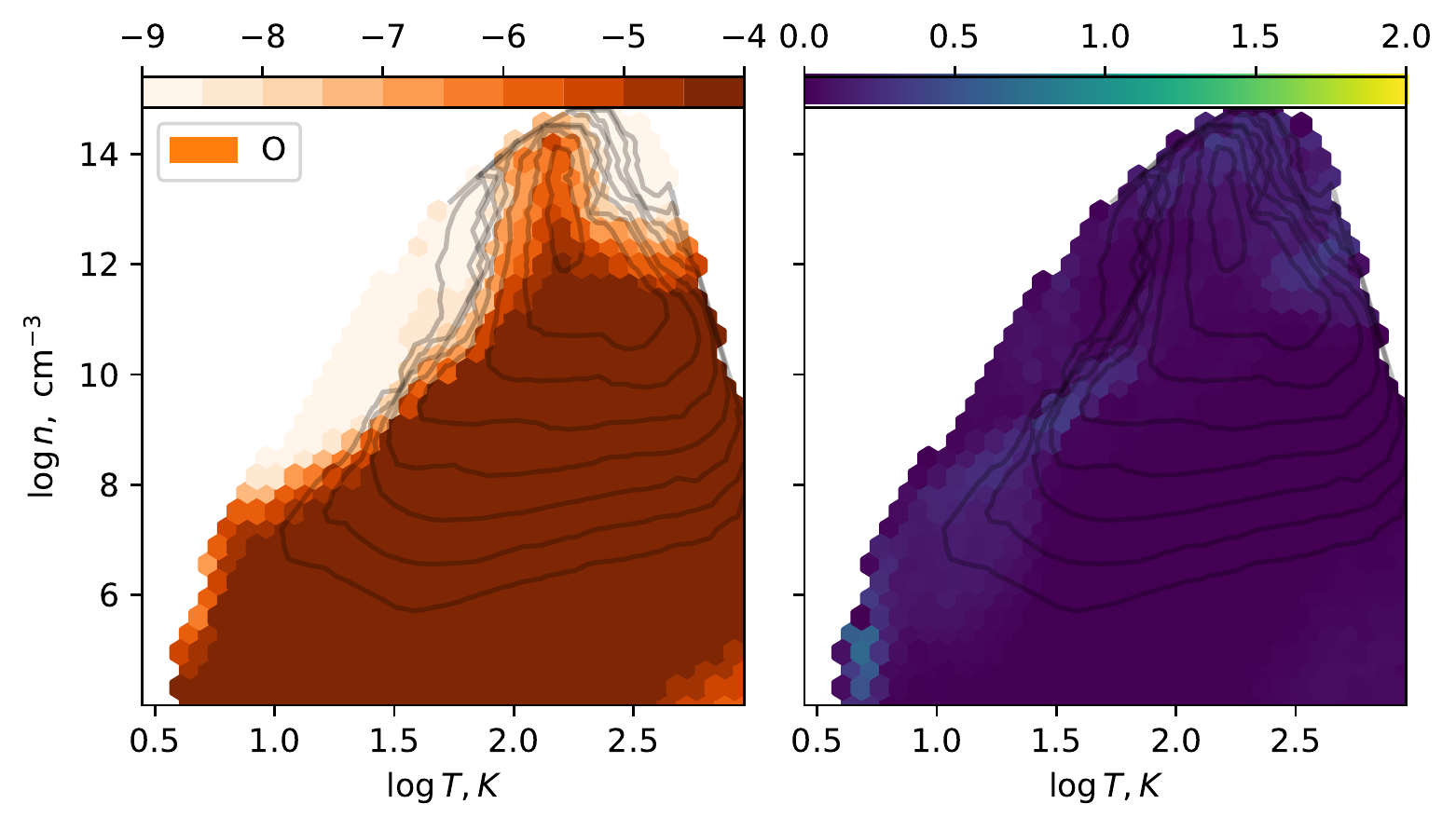}
    \includegraphics[width=0.37\linewidth]{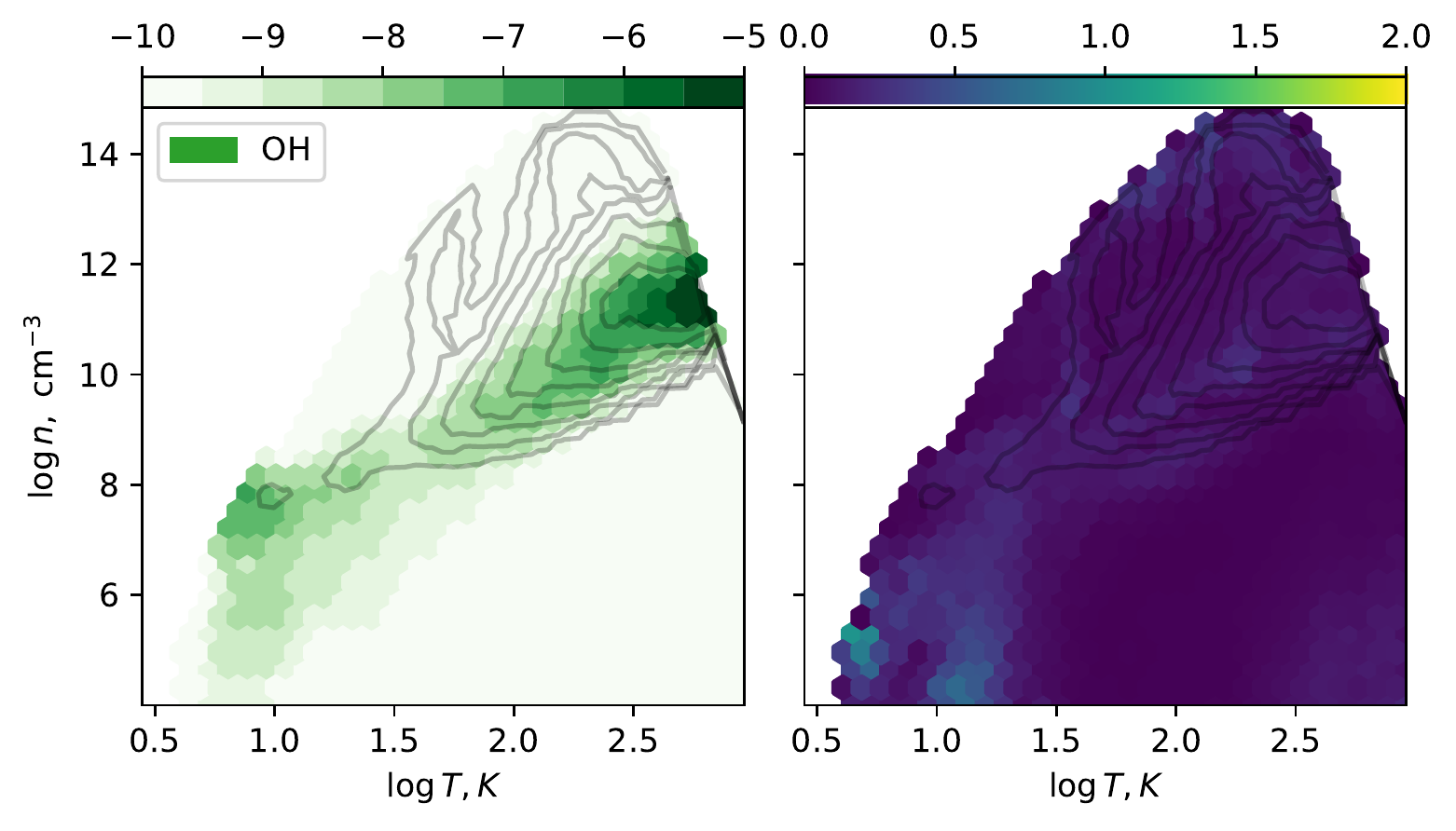}

    \includegraphics[width=0.37\linewidth]{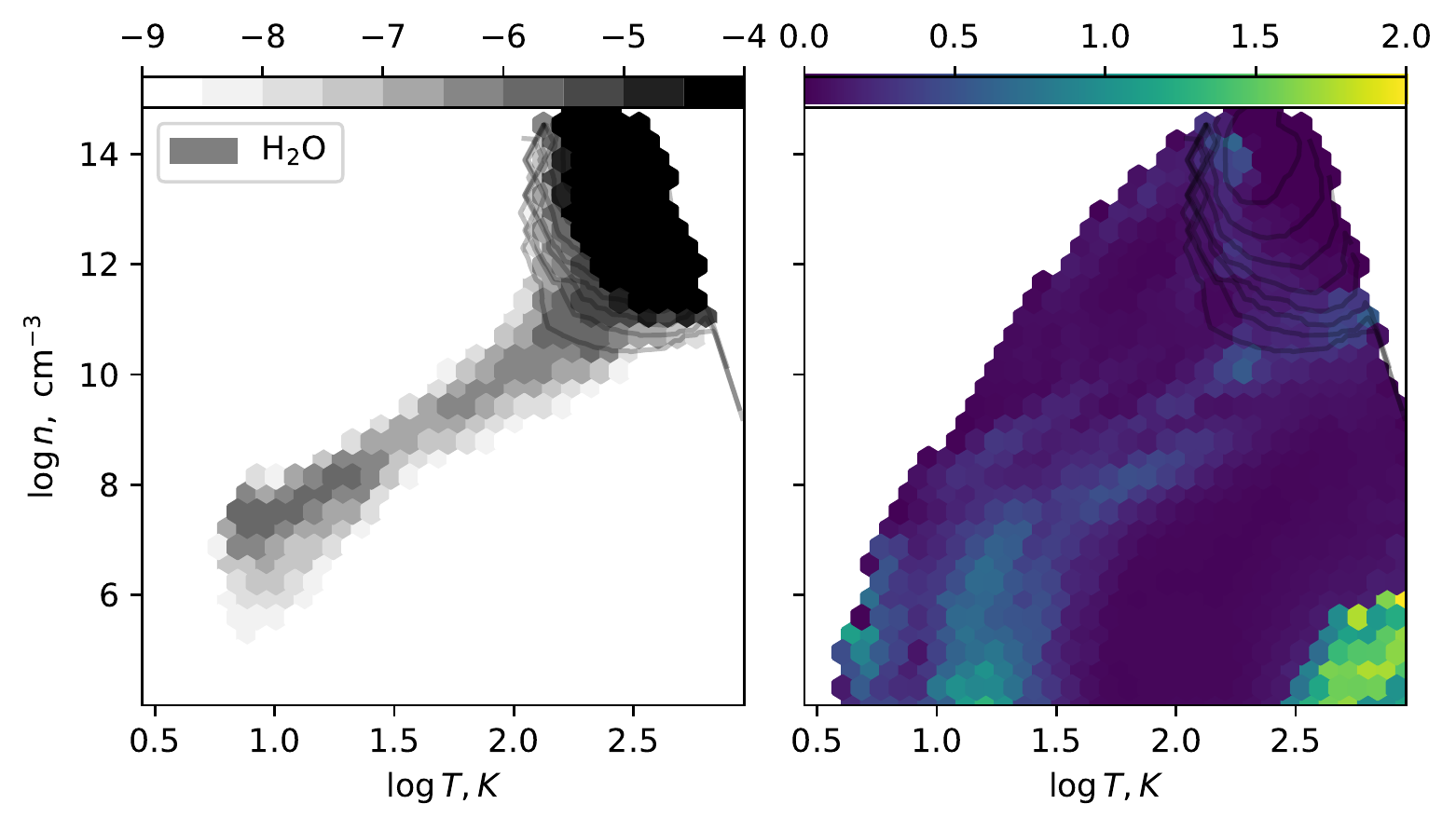}
    \includegraphics[width=0.37\linewidth]{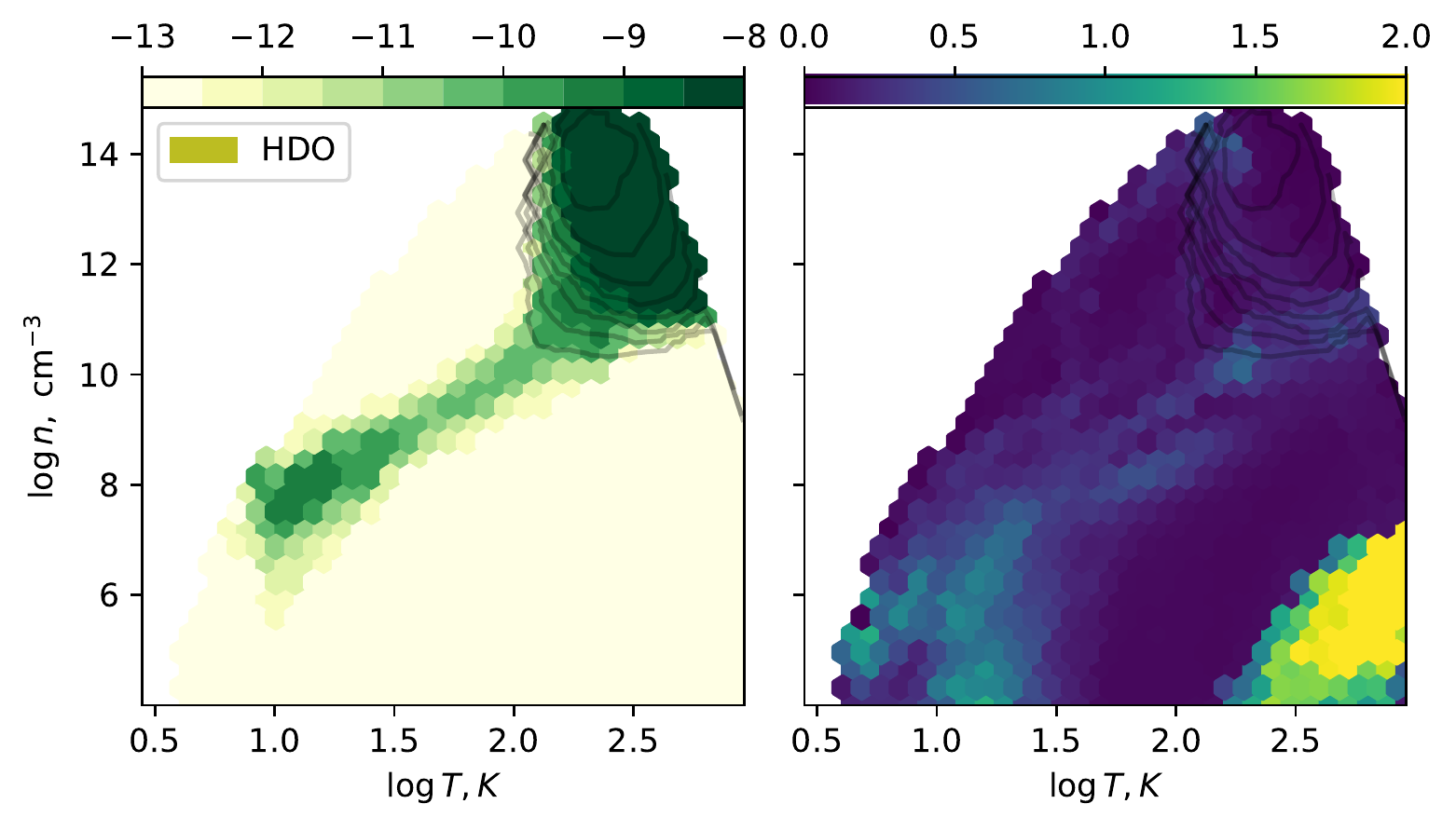}

    \includegraphics[width=0.37\linewidth]{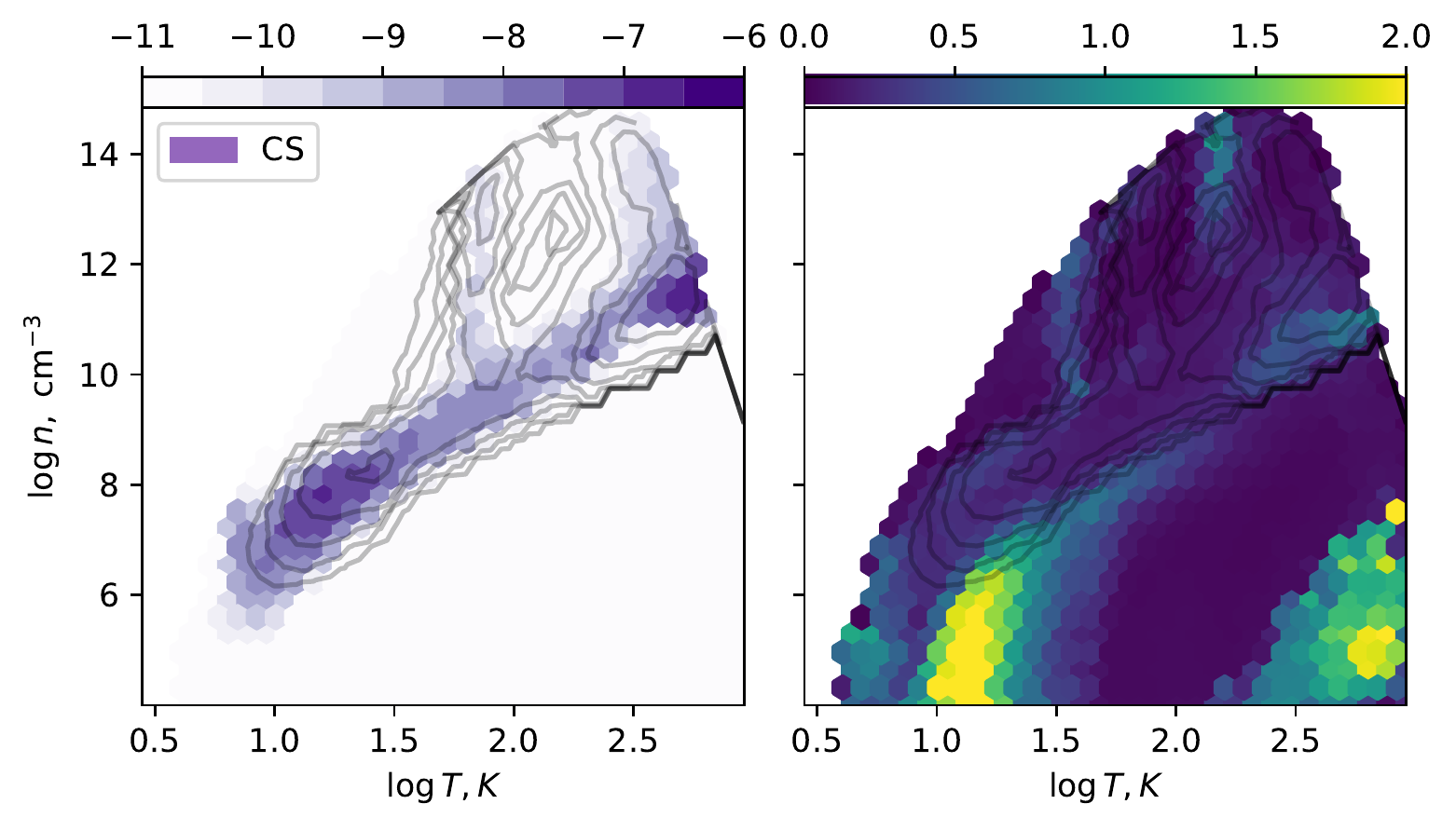}
    \includegraphics[width=0.37\linewidth]{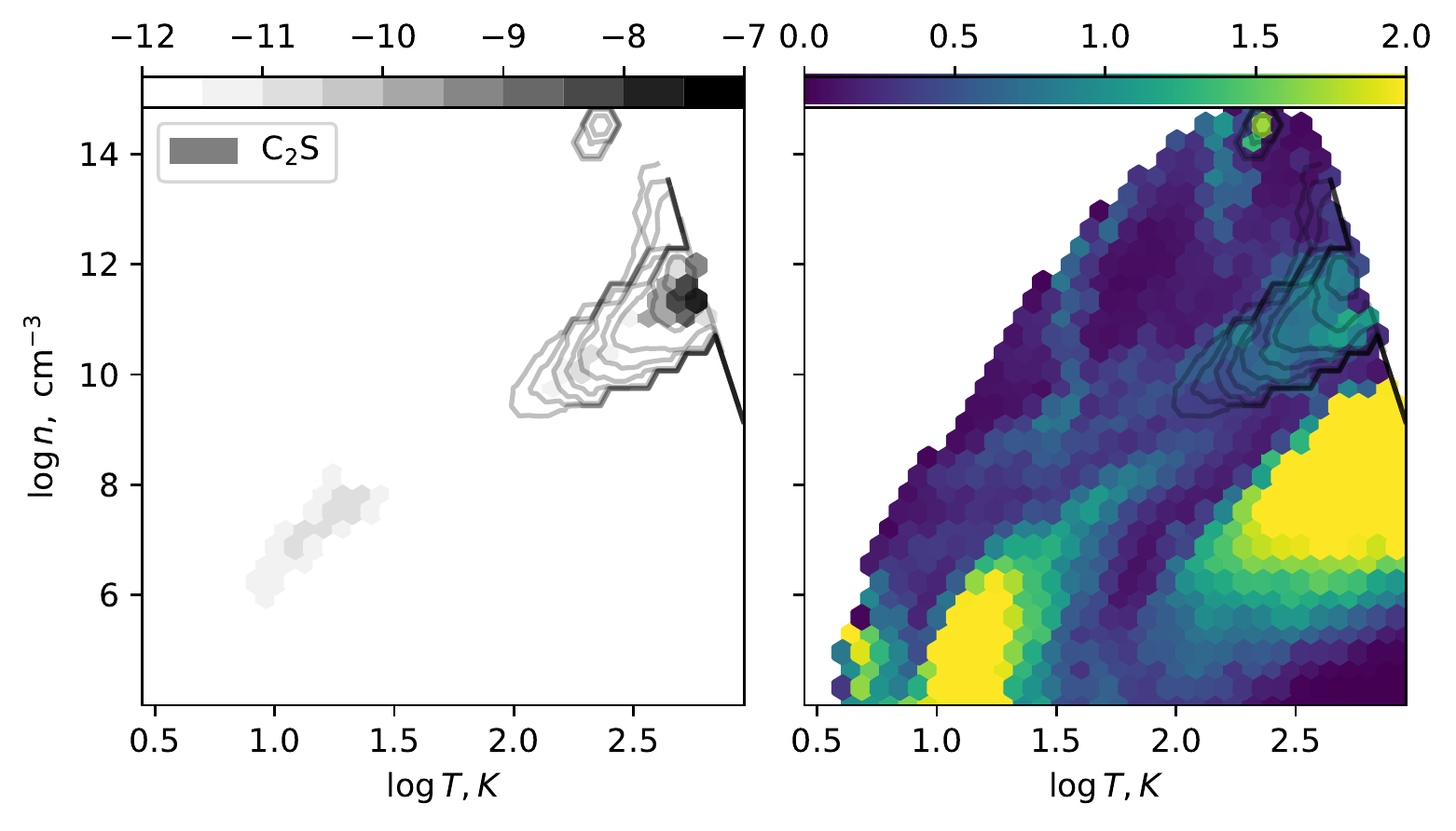}

    \includegraphics[width=0.37\linewidth]{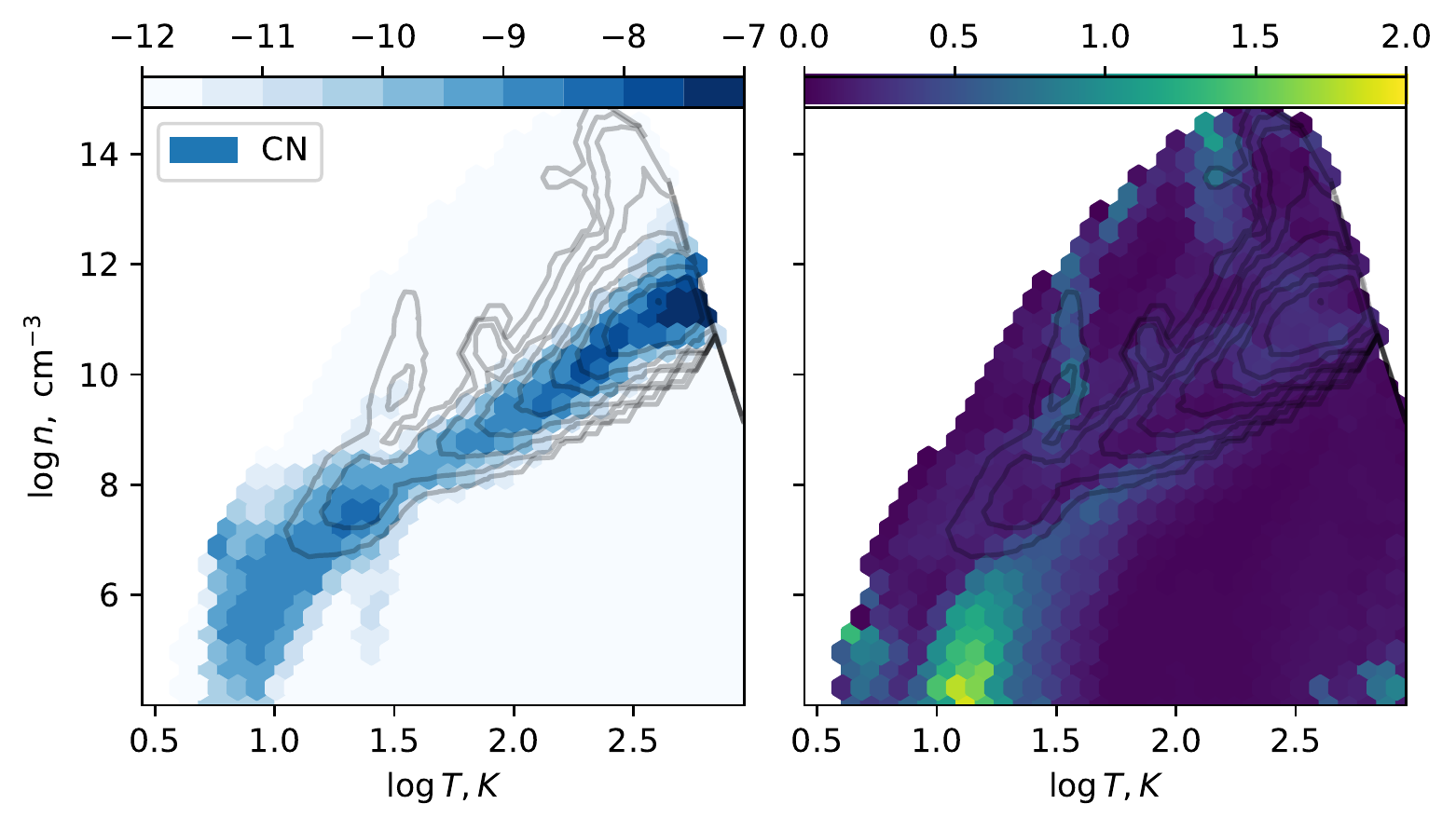}
    \includegraphics[width=0.37\linewidth]{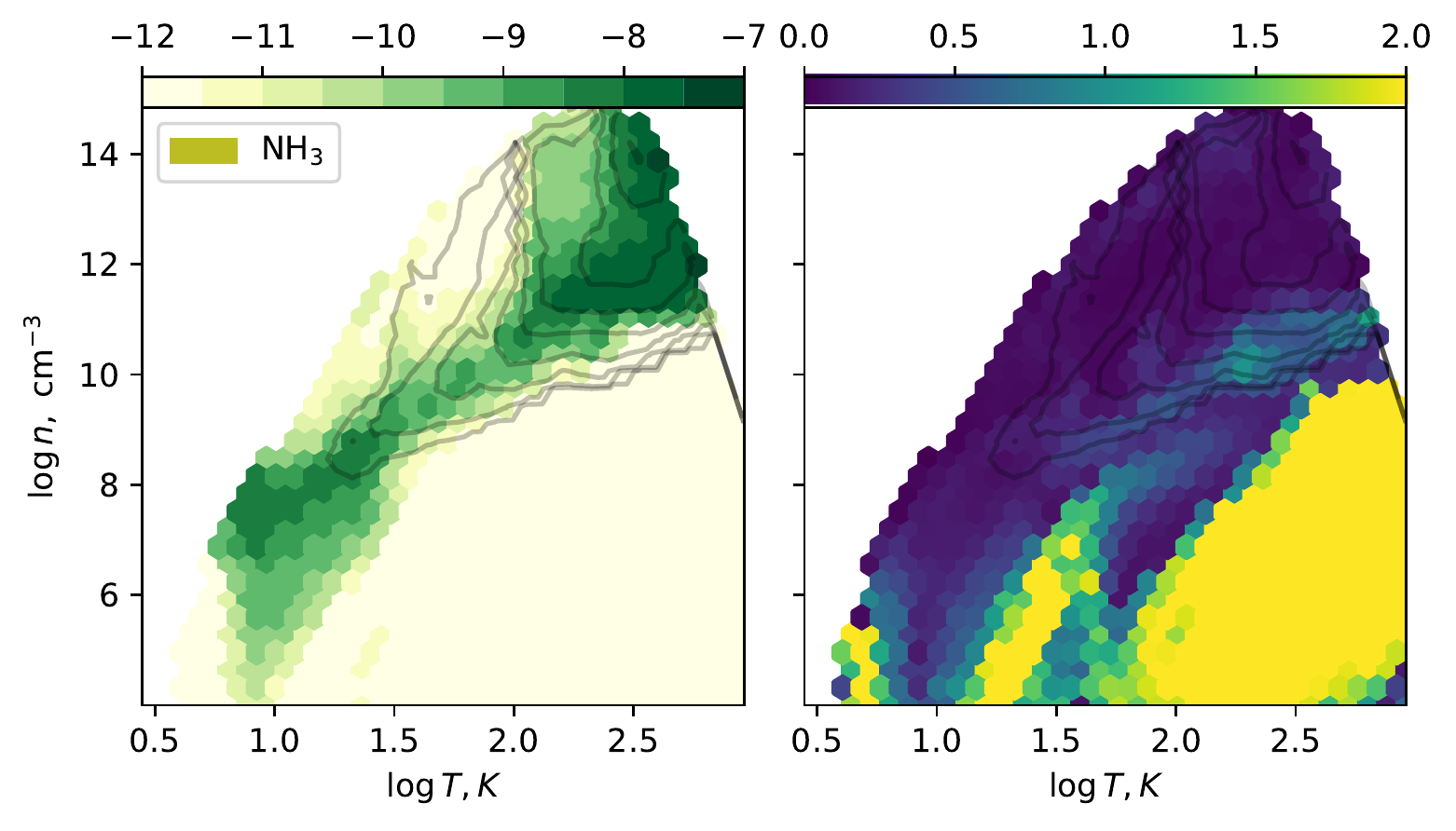}

    \includegraphics[width=0.37\linewidth]{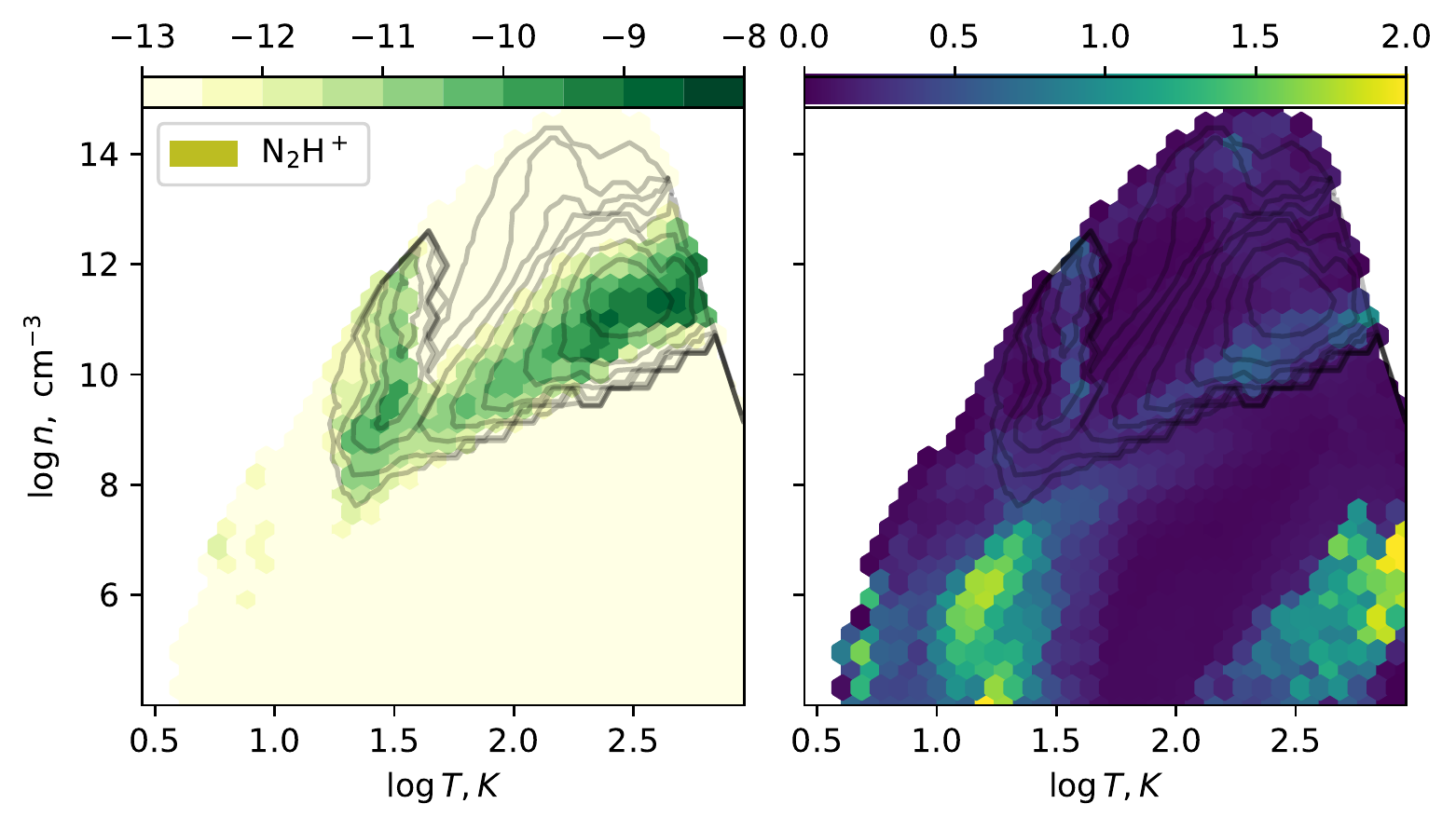}
    \includegraphics[width=0.37\linewidth]{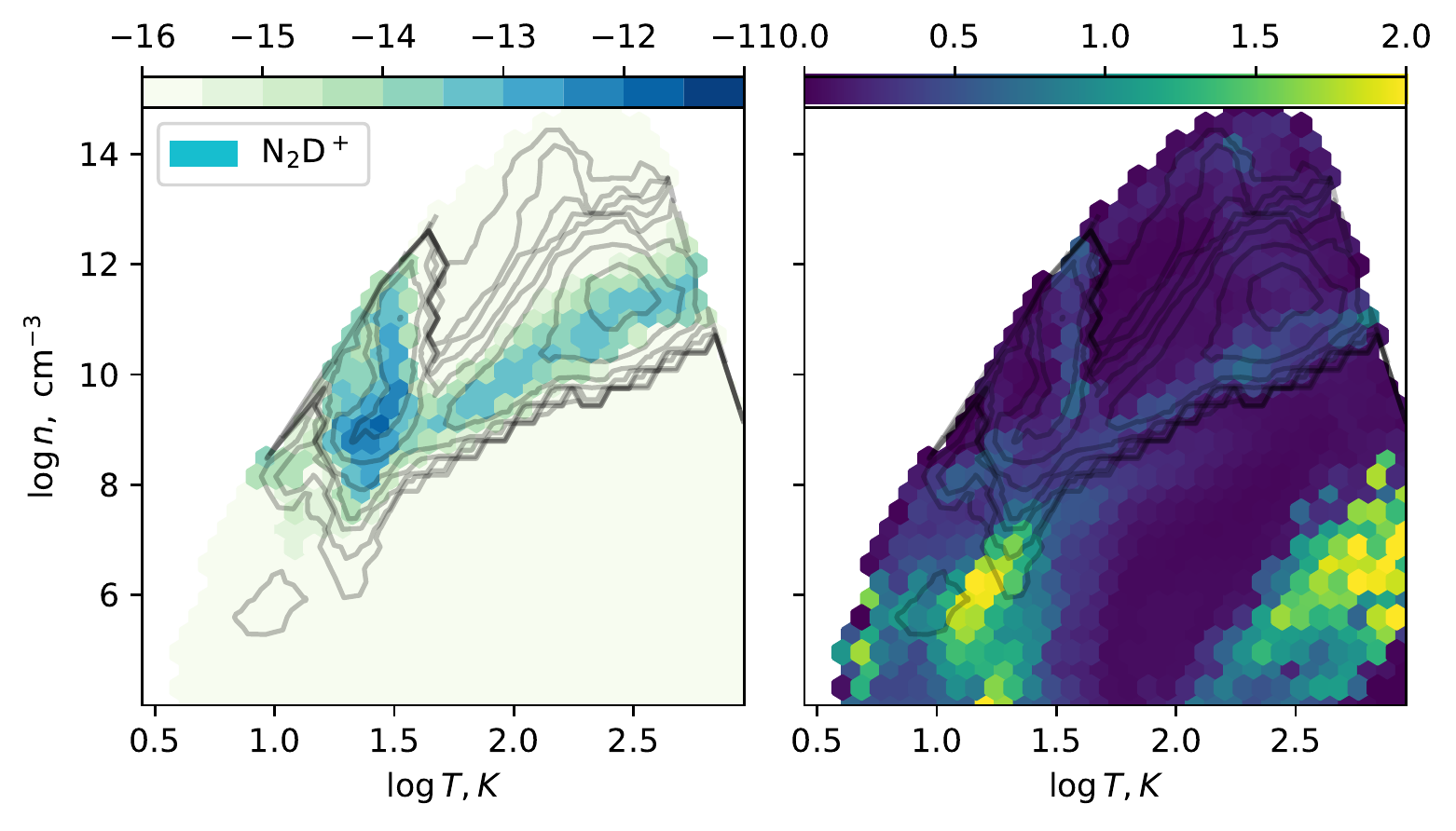}

    \caption{Application of the same method to a broader set of species. Performance of ML-accelerated chemistry prediction, same as in Fig.~\ref{fig:performance}, based on four input features.}
    \label{fig:morespecies}

\end{figure*}

\setcounter{figure}{2}
\begin{figure*}
\centering
    \includegraphics[width=0.37\linewidth]{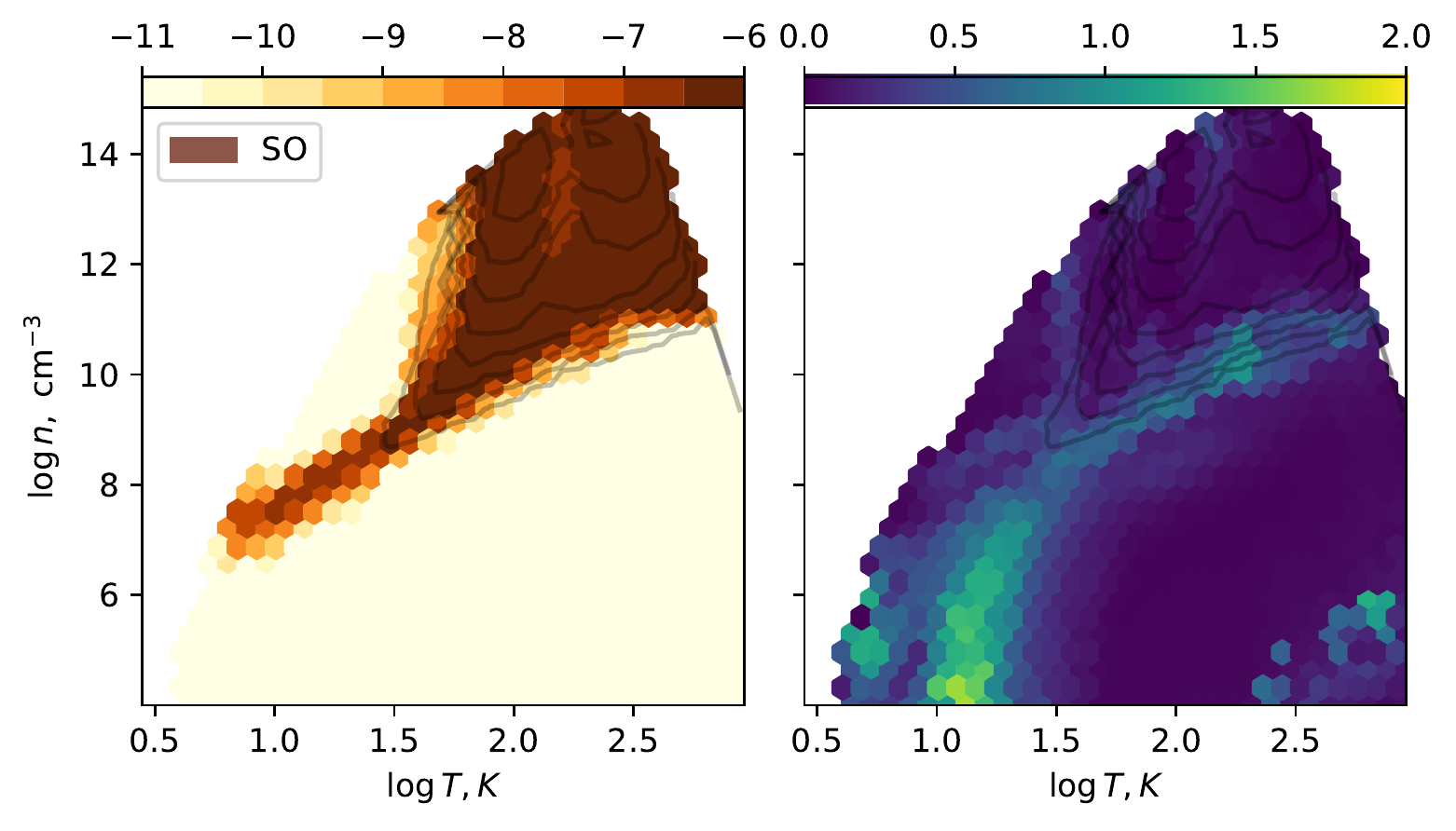}
    \includegraphics[width=0.37\linewidth]{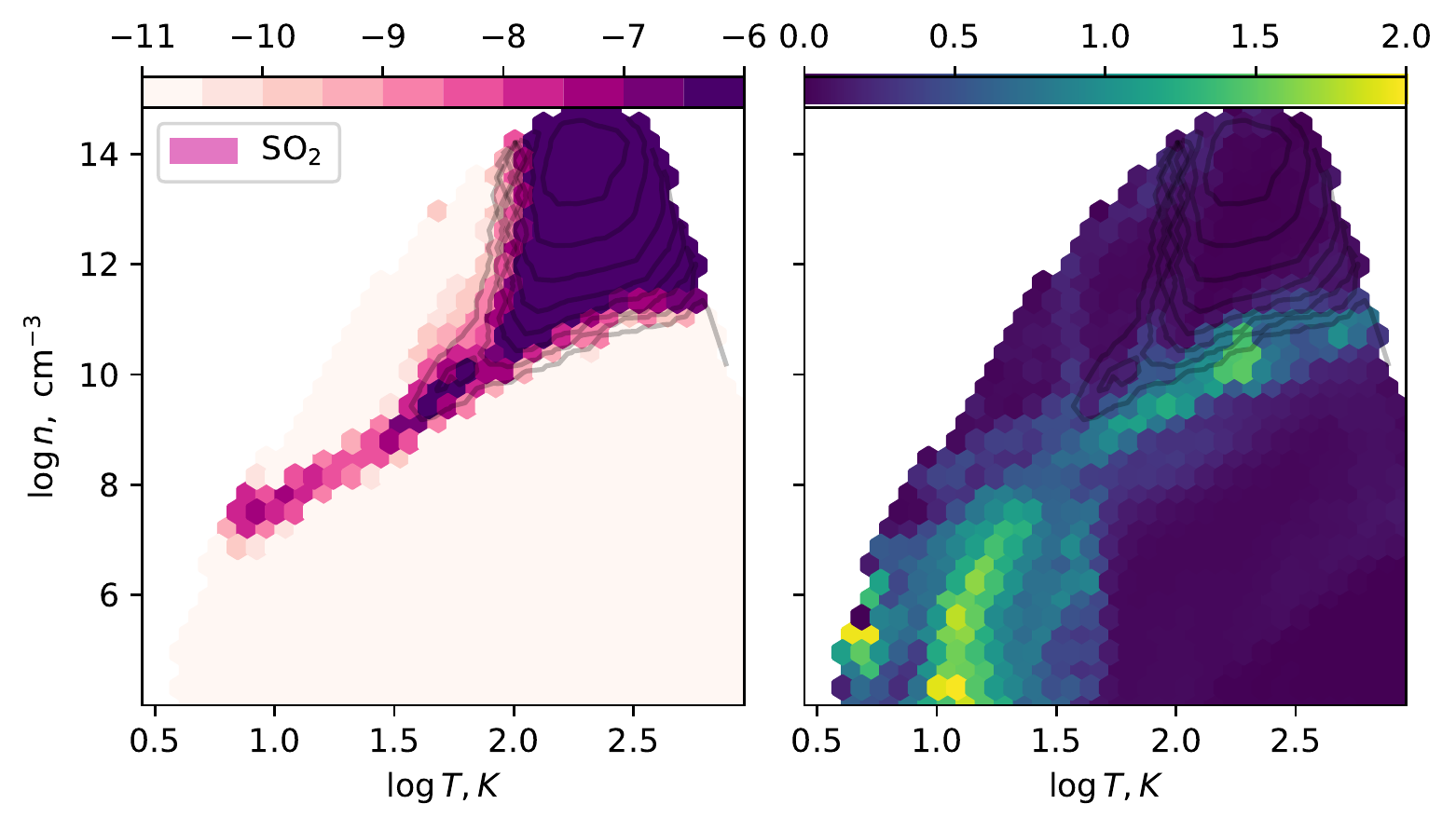}

    \includegraphics[width=0.37\linewidth]{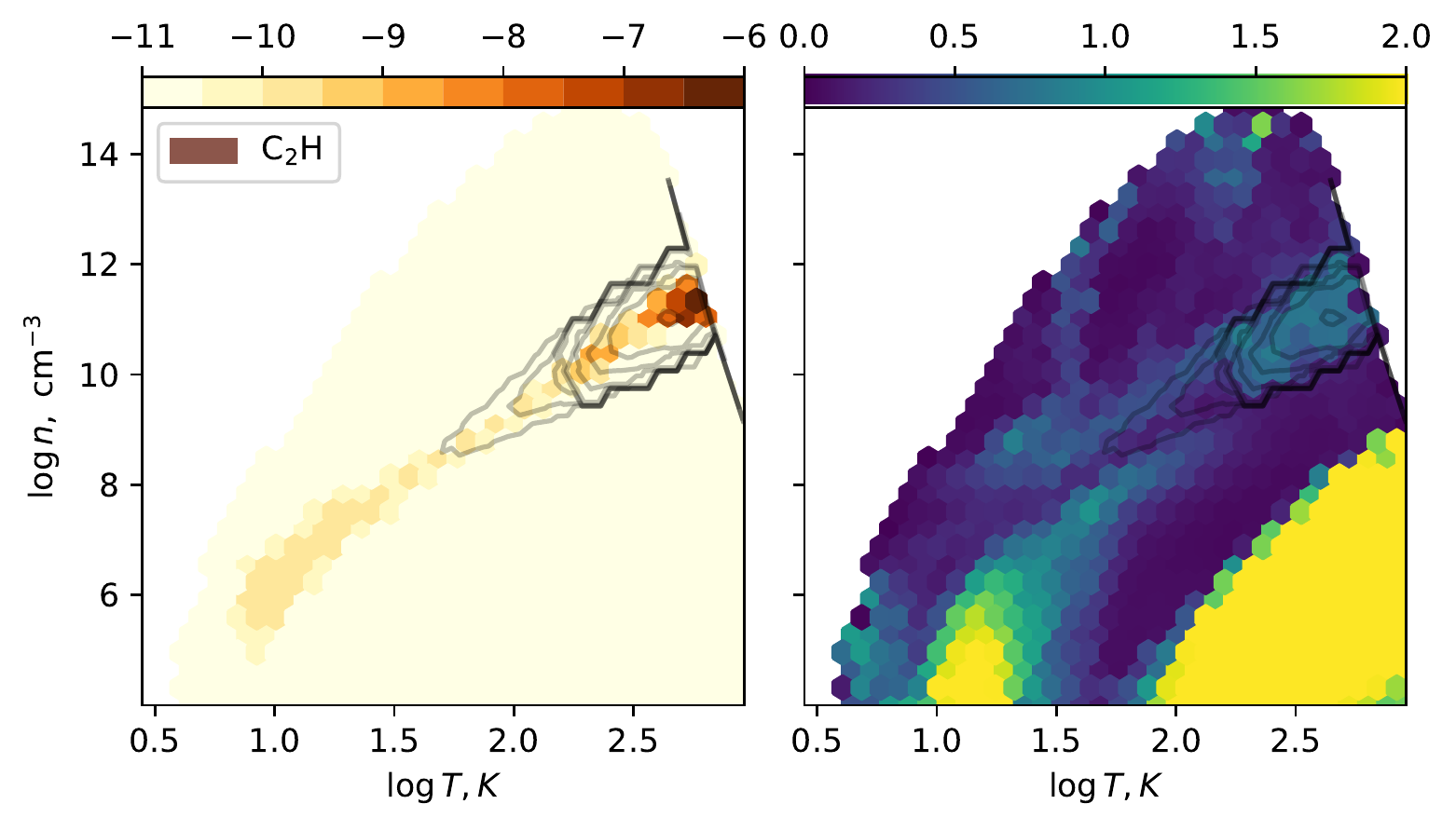}
    \includegraphics[width=0.37\linewidth]{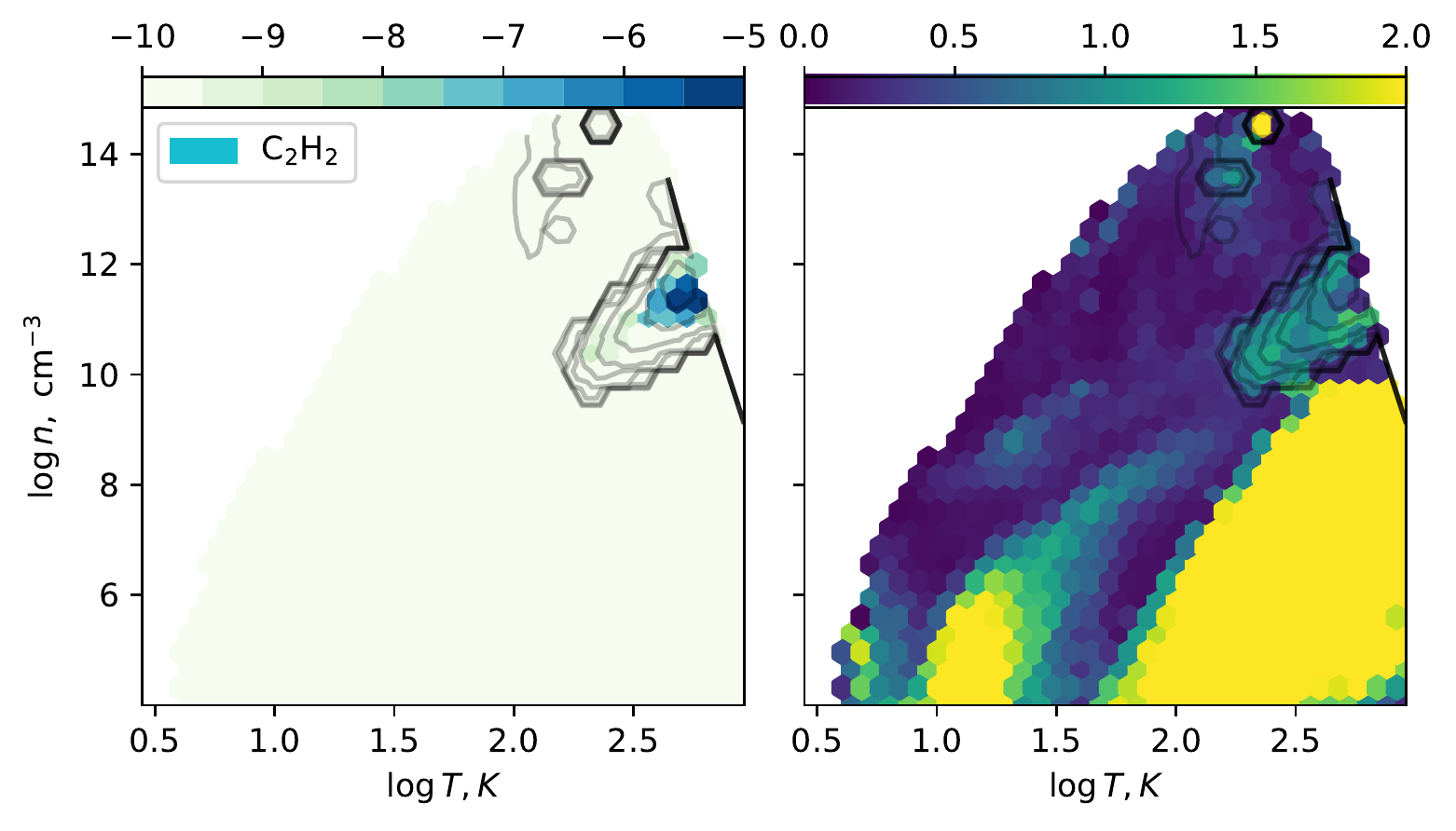}

    \includegraphics[width=0.37\linewidth]{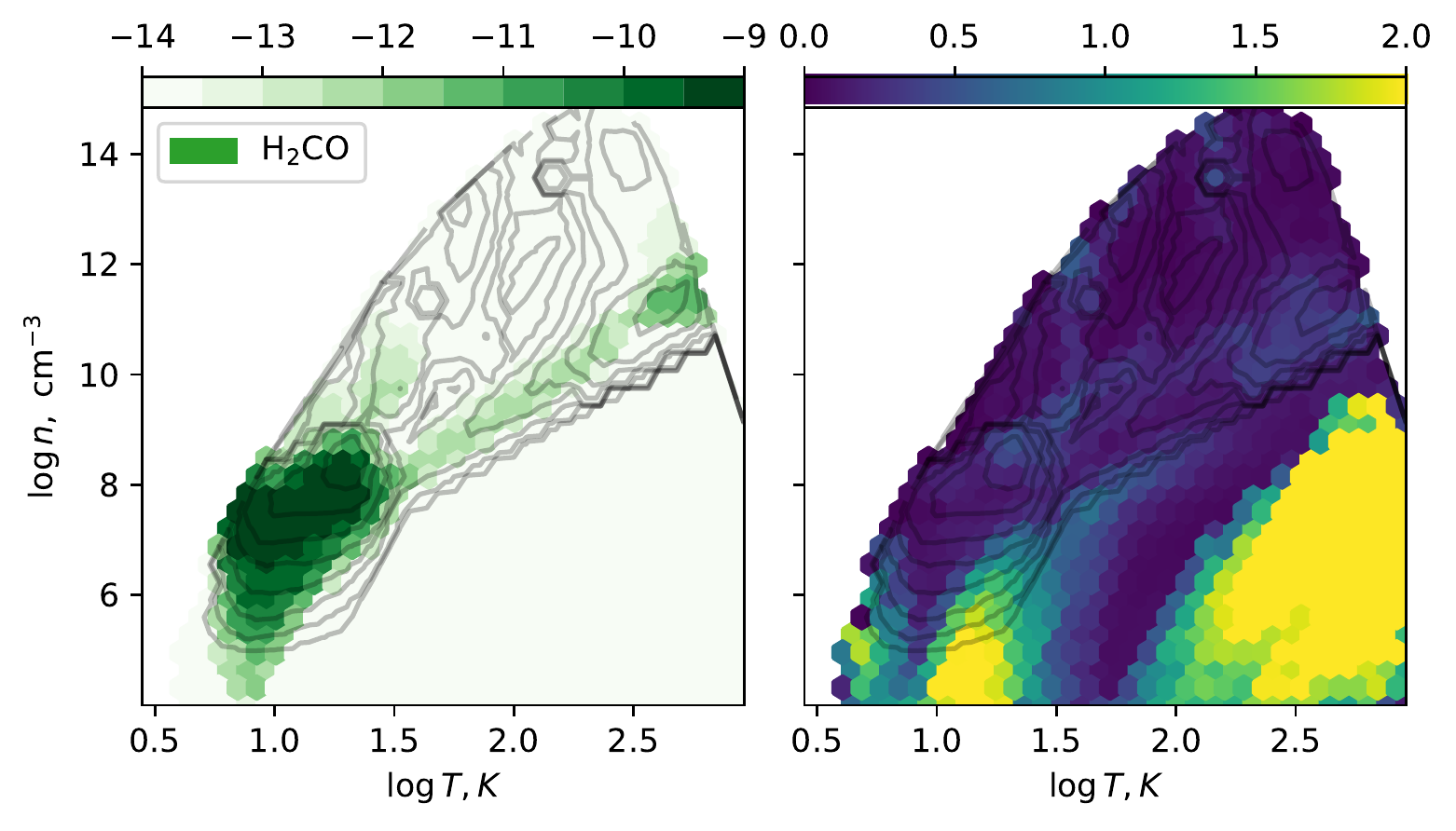}
    \includegraphics[width=0.37\linewidth]{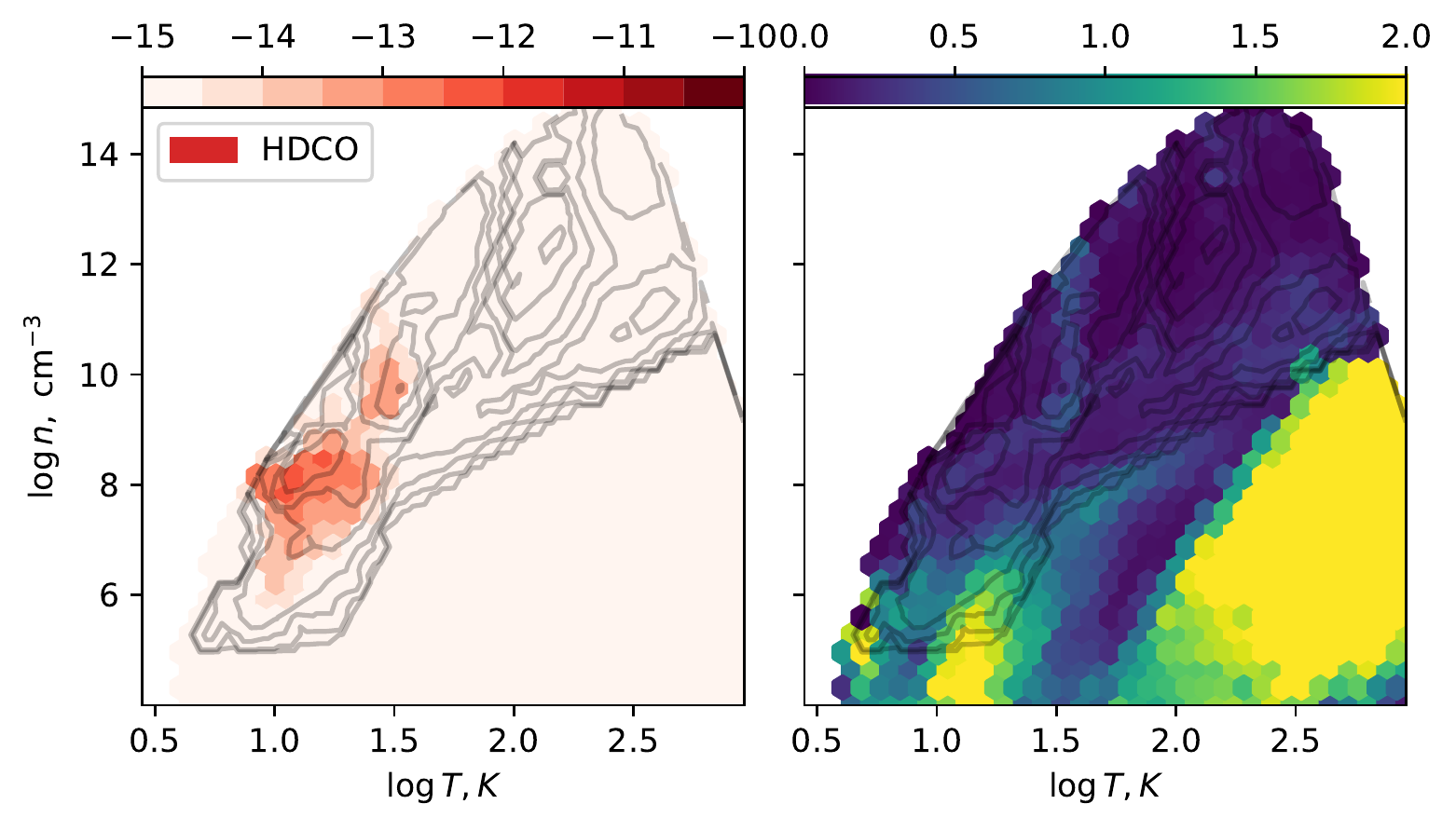}

    \includegraphics[width=0.37\linewidth]{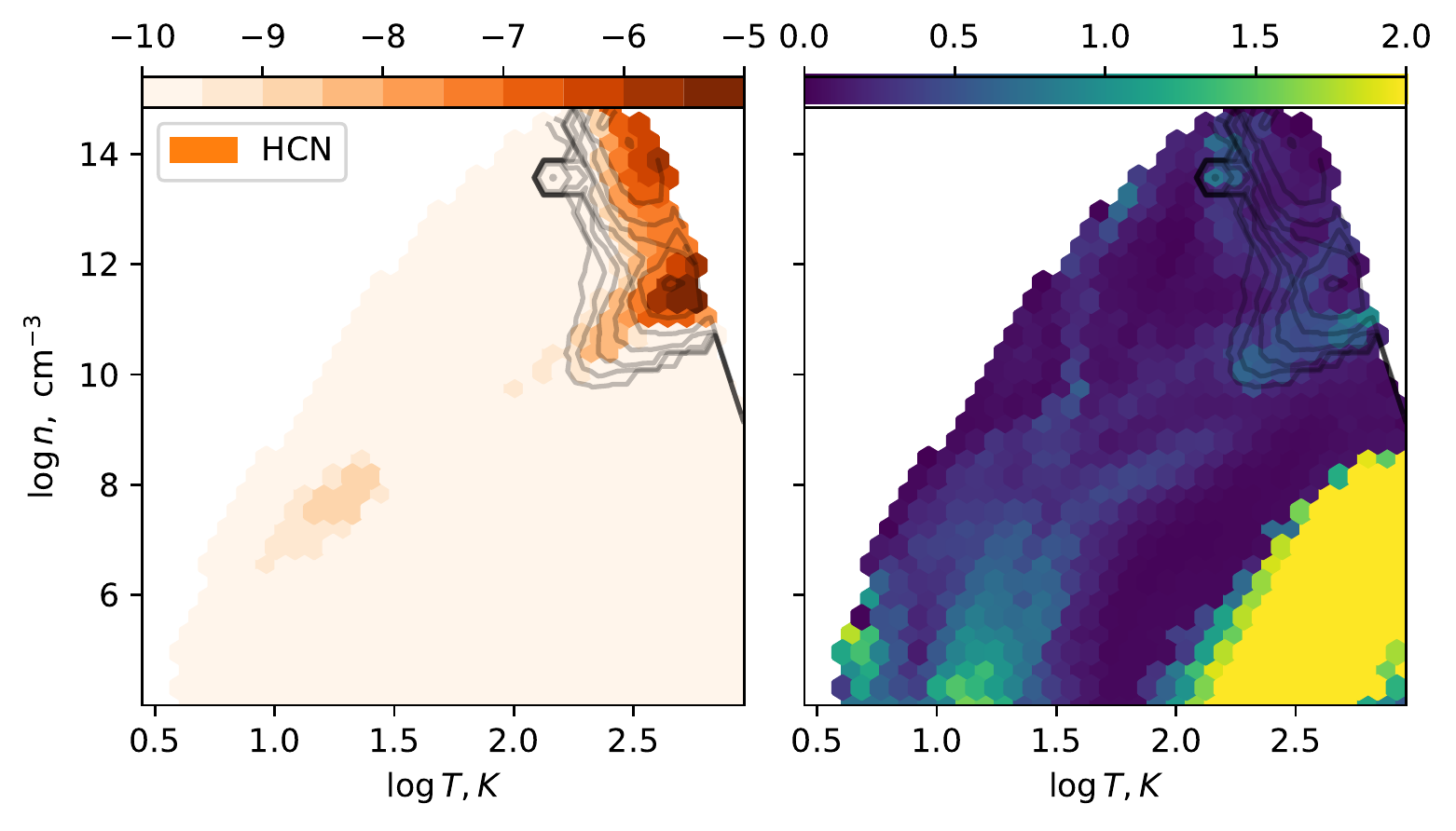}
    \includegraphics[width=0.37\linewidth]{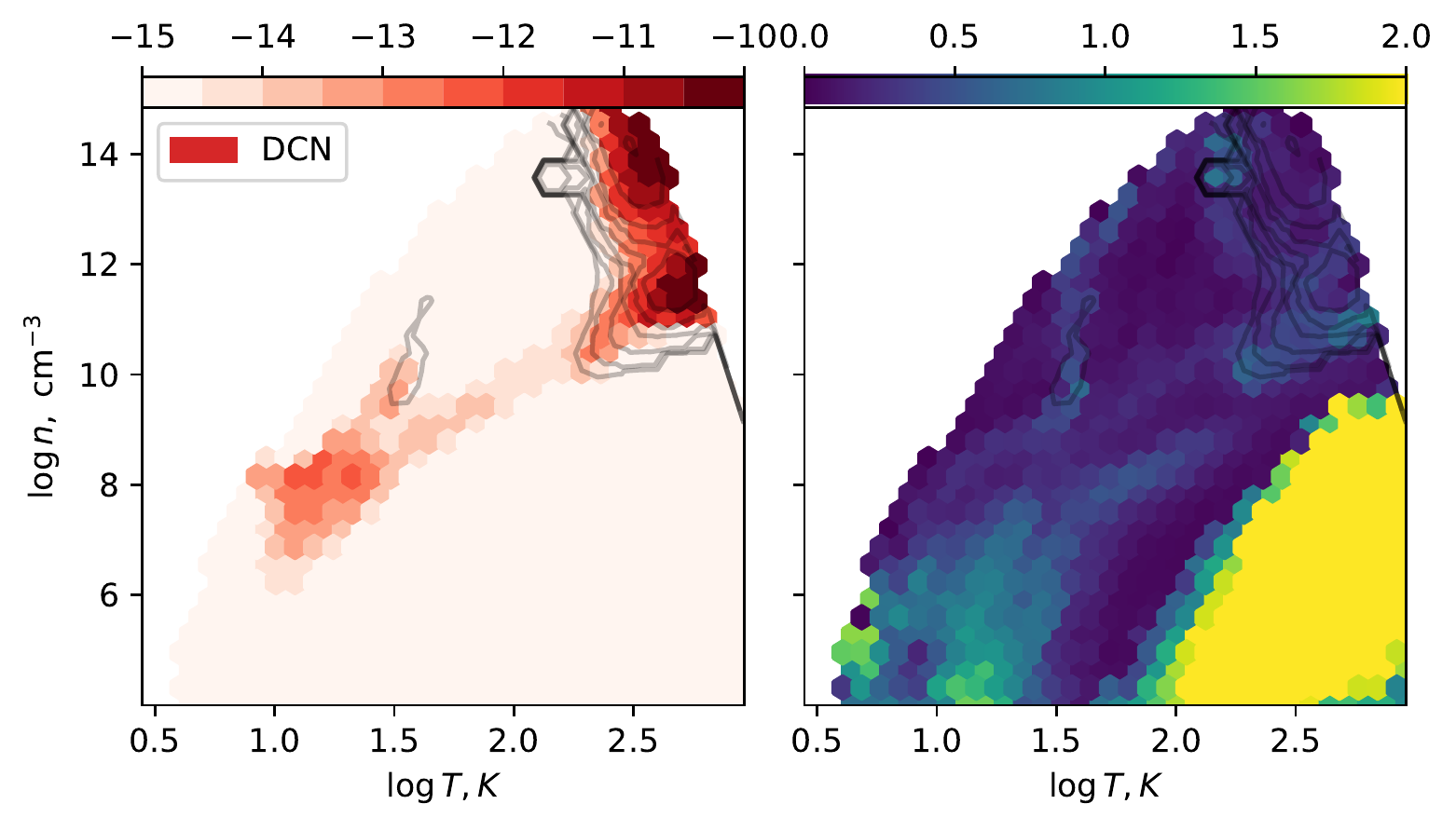}

    \includegraphics[width=0.37\linewidth]{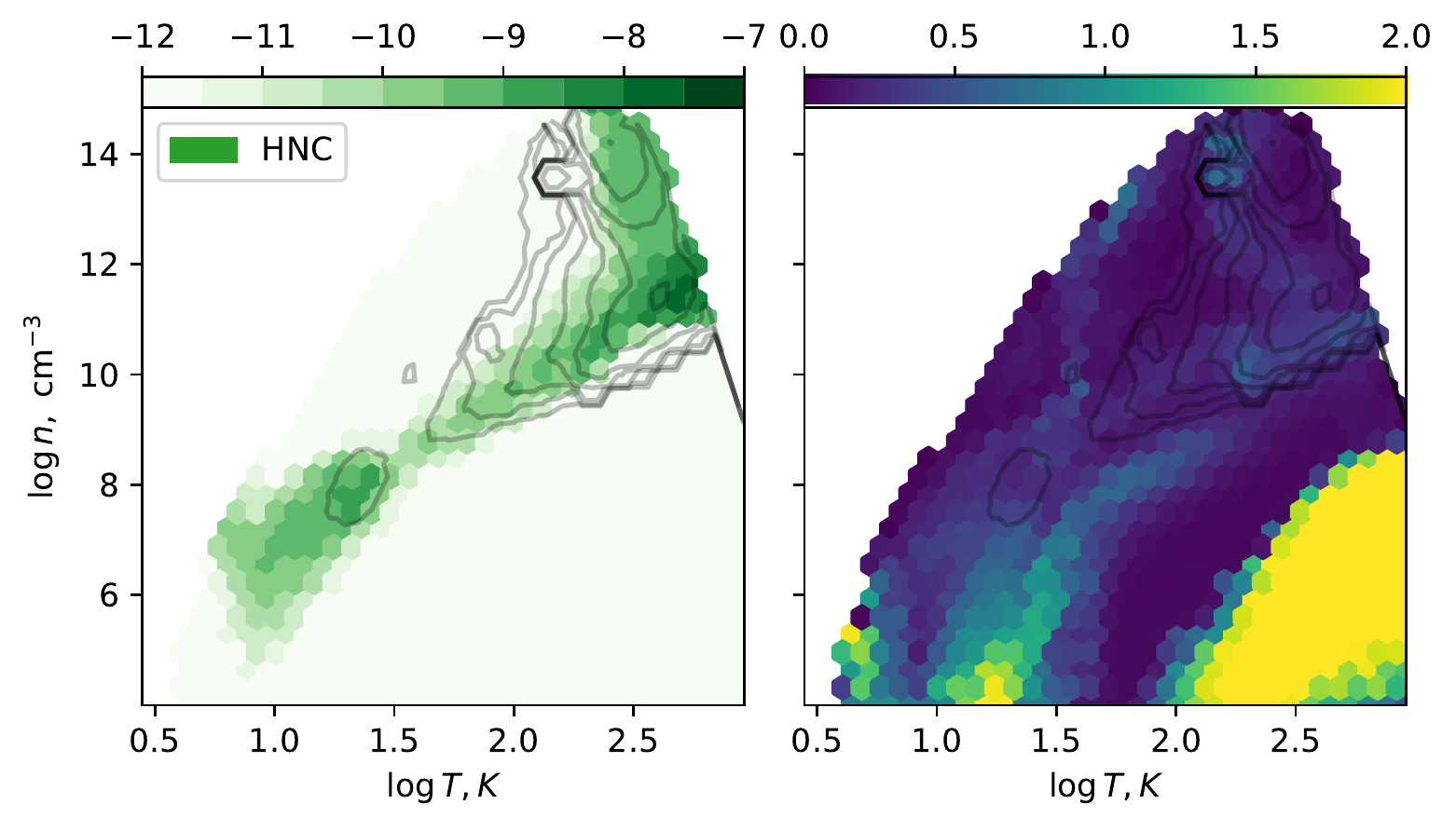}
    \includegraphics[width=0.37\linewidth]{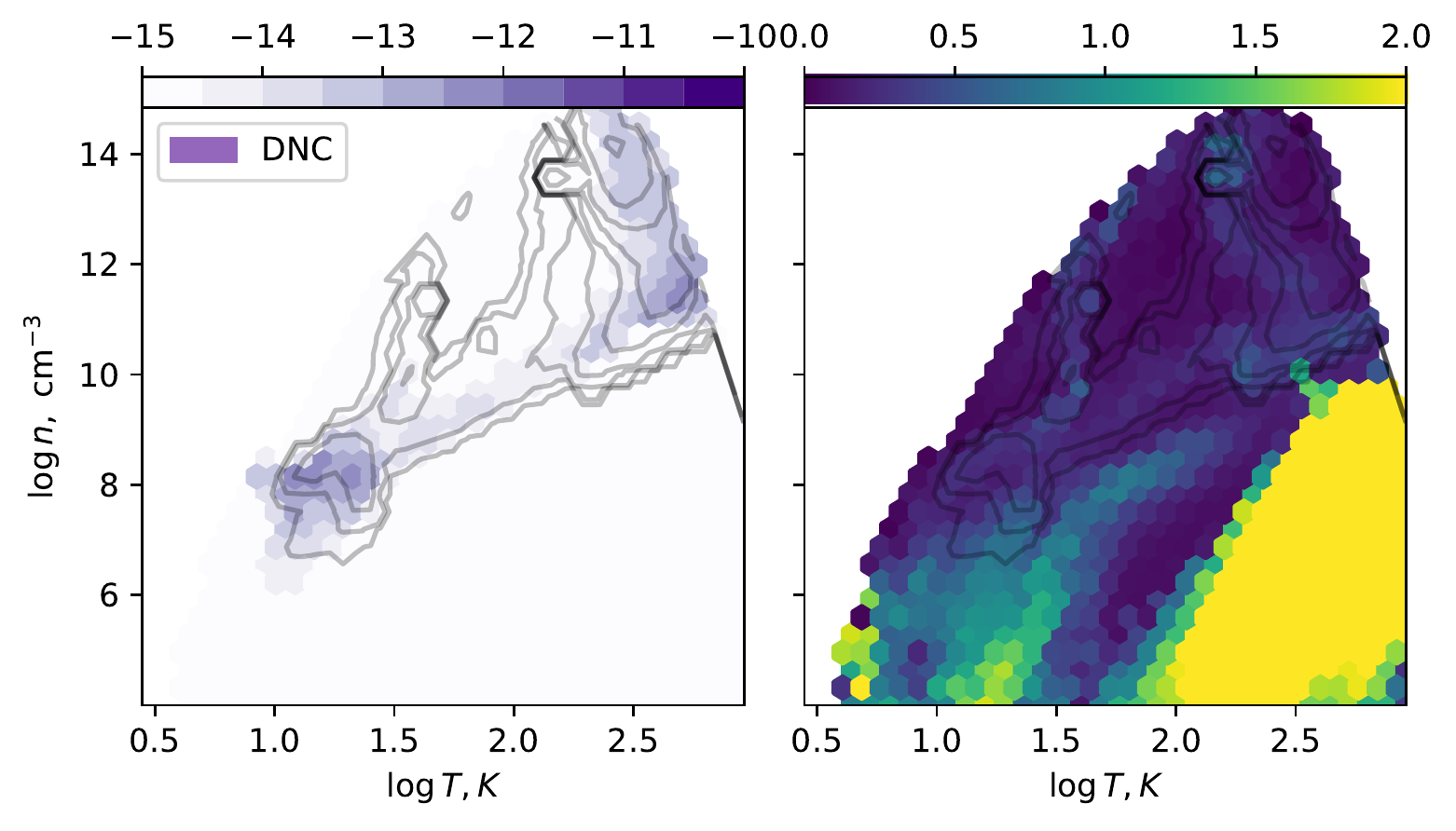}

    \includegraphics[width=0.37\linewidth]{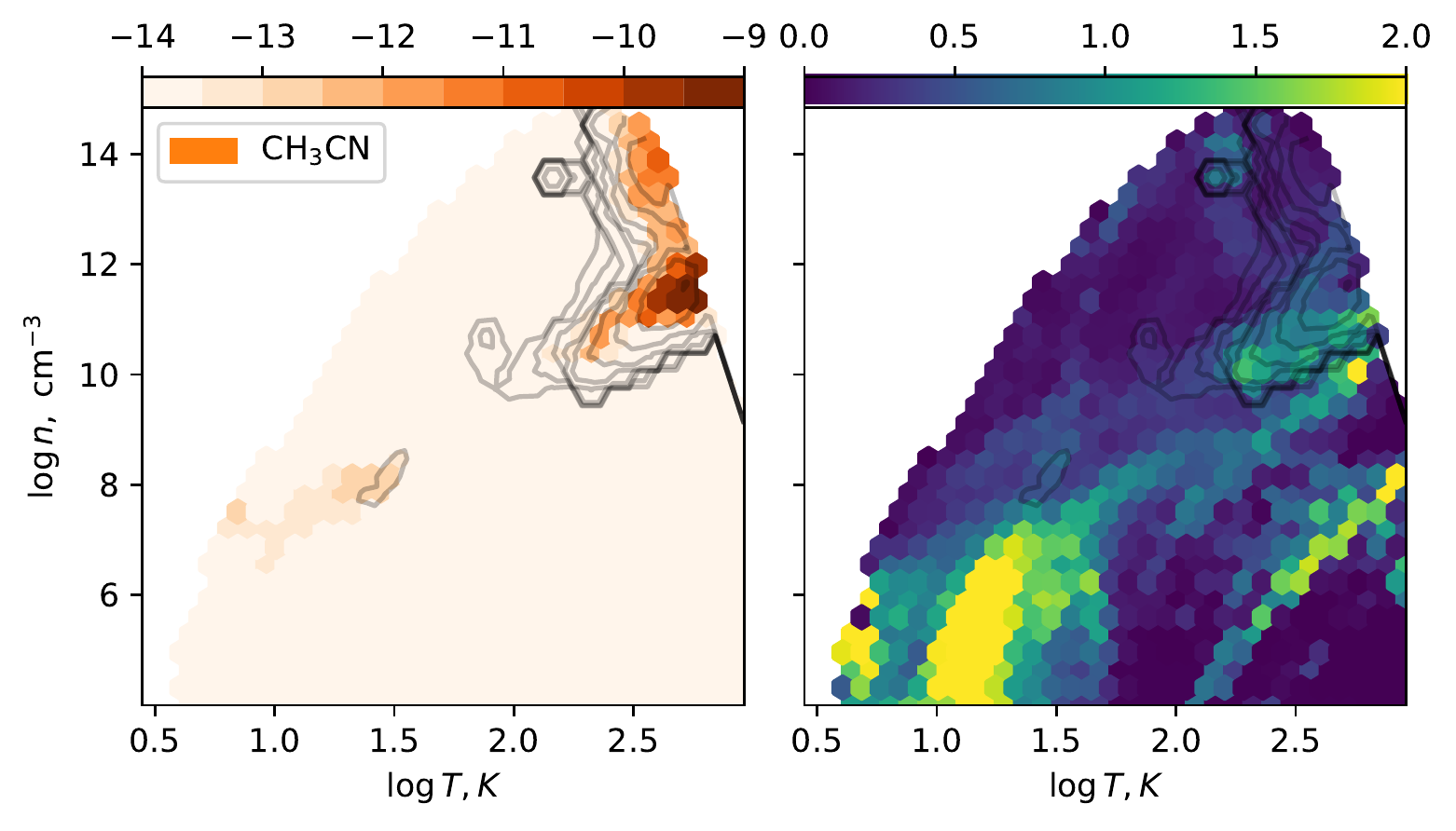}
    \includegraphics[width=0.37\linewidth]{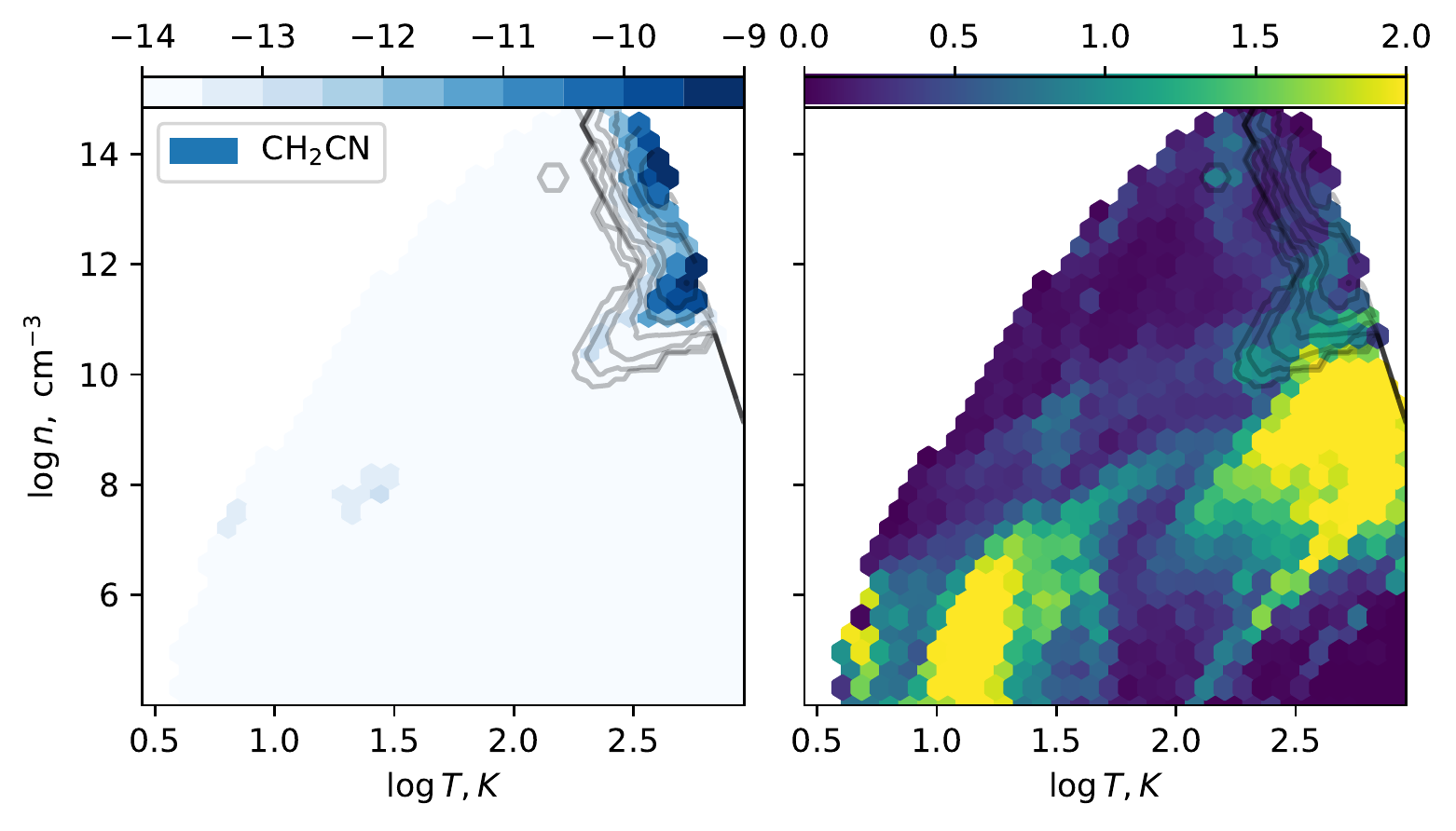}

    \caption{continued.}
    \label{fig:morespecies2}
\end{figure*}

\setcounter{figure}{2}
\begin{figure*}
\centering
    \includegraphics[width=0.37\linewidth]{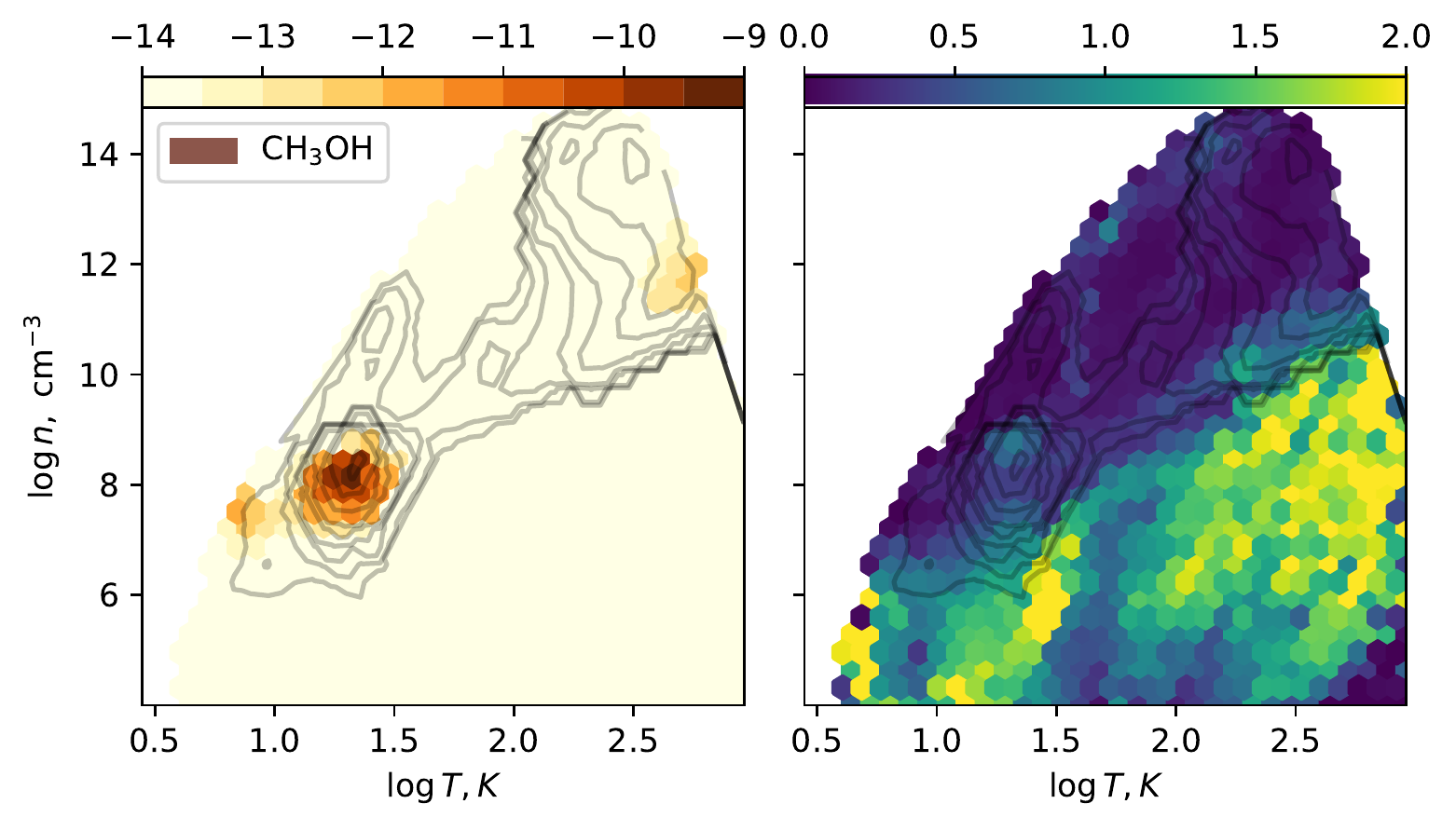}
    \includegraphics[width=0.37\linewidth]{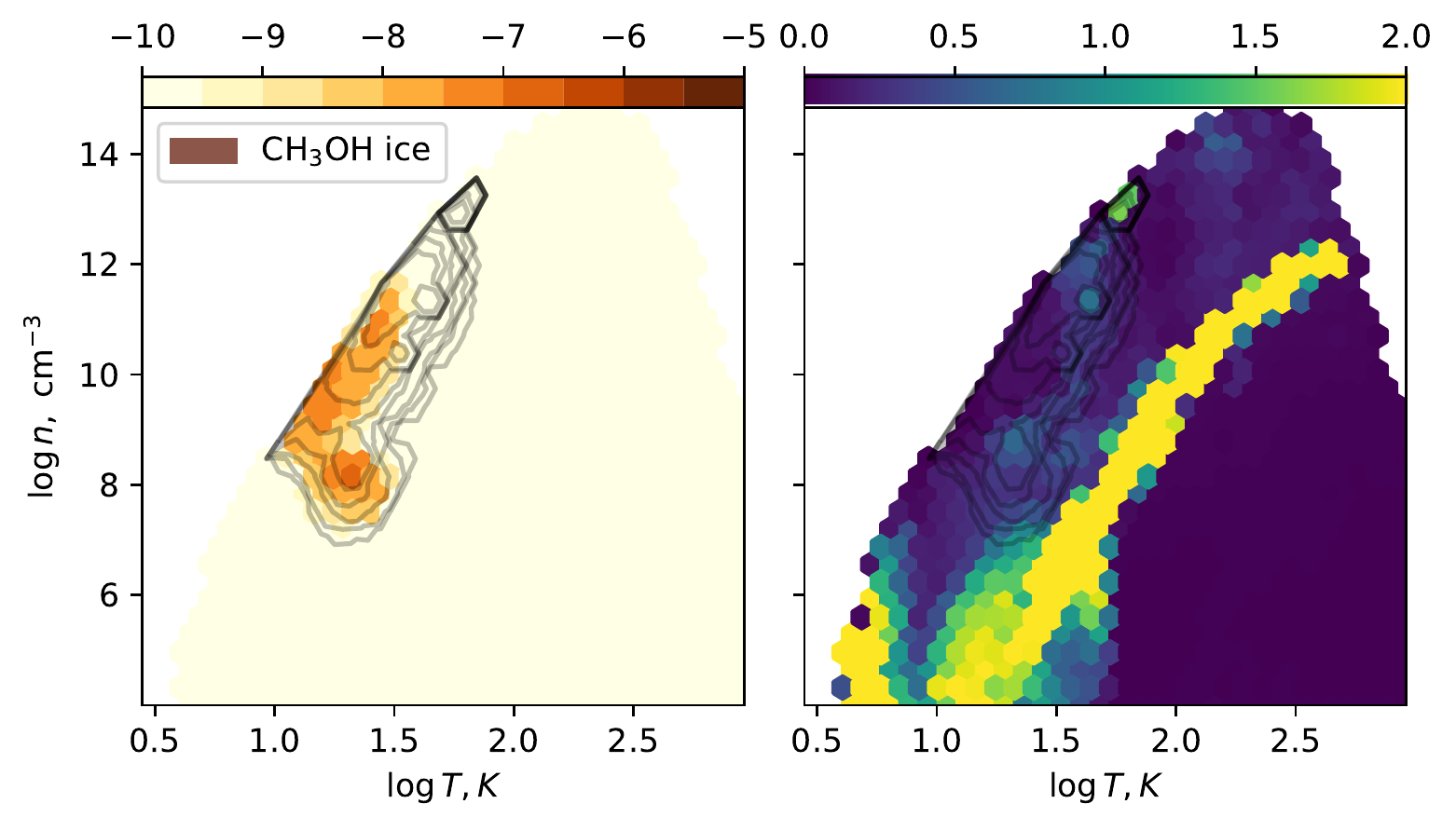}

    \includegraphics[width=0.37\linewidth]{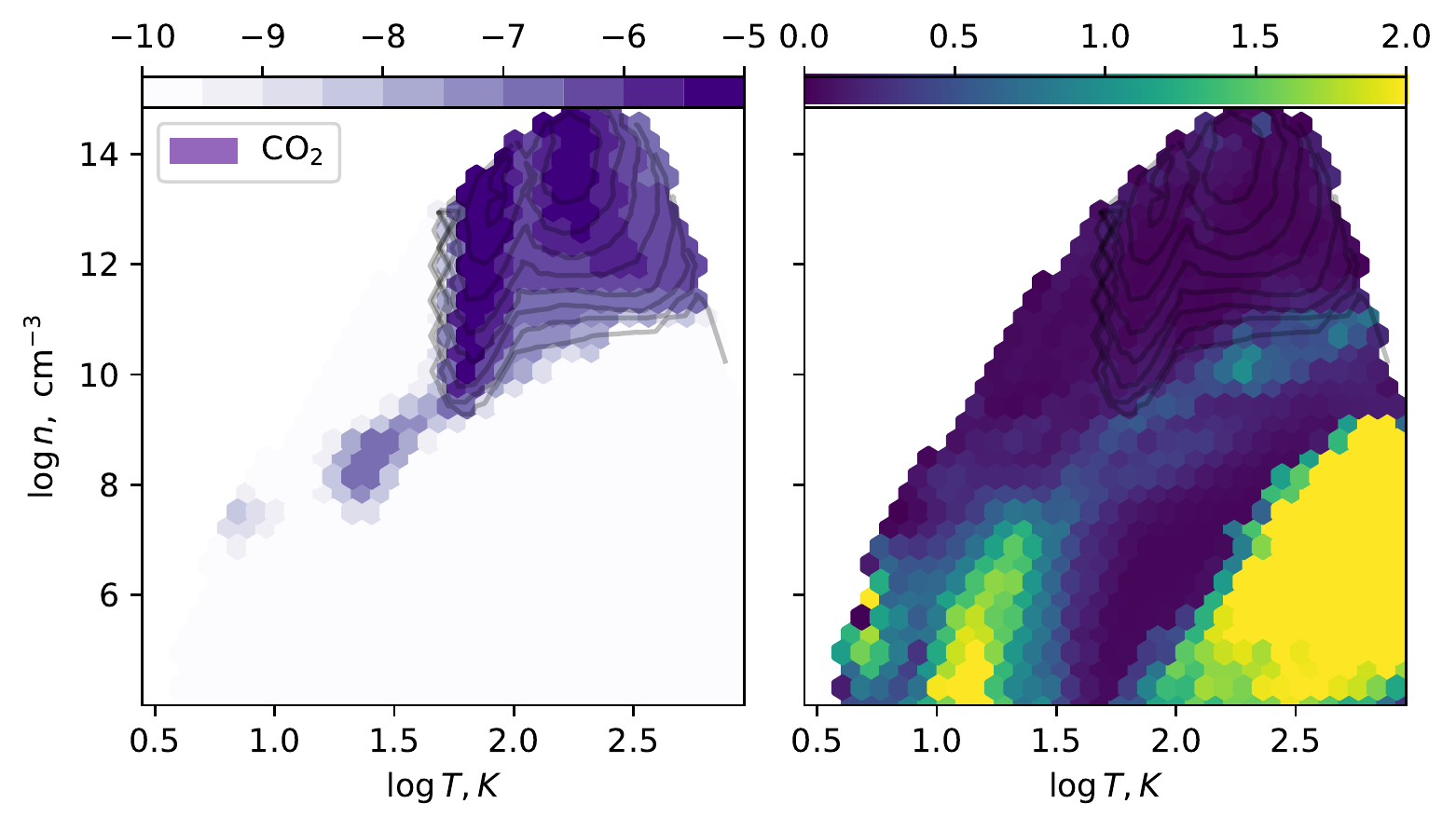}
    \includegraphics[width=0.37\linewidth]{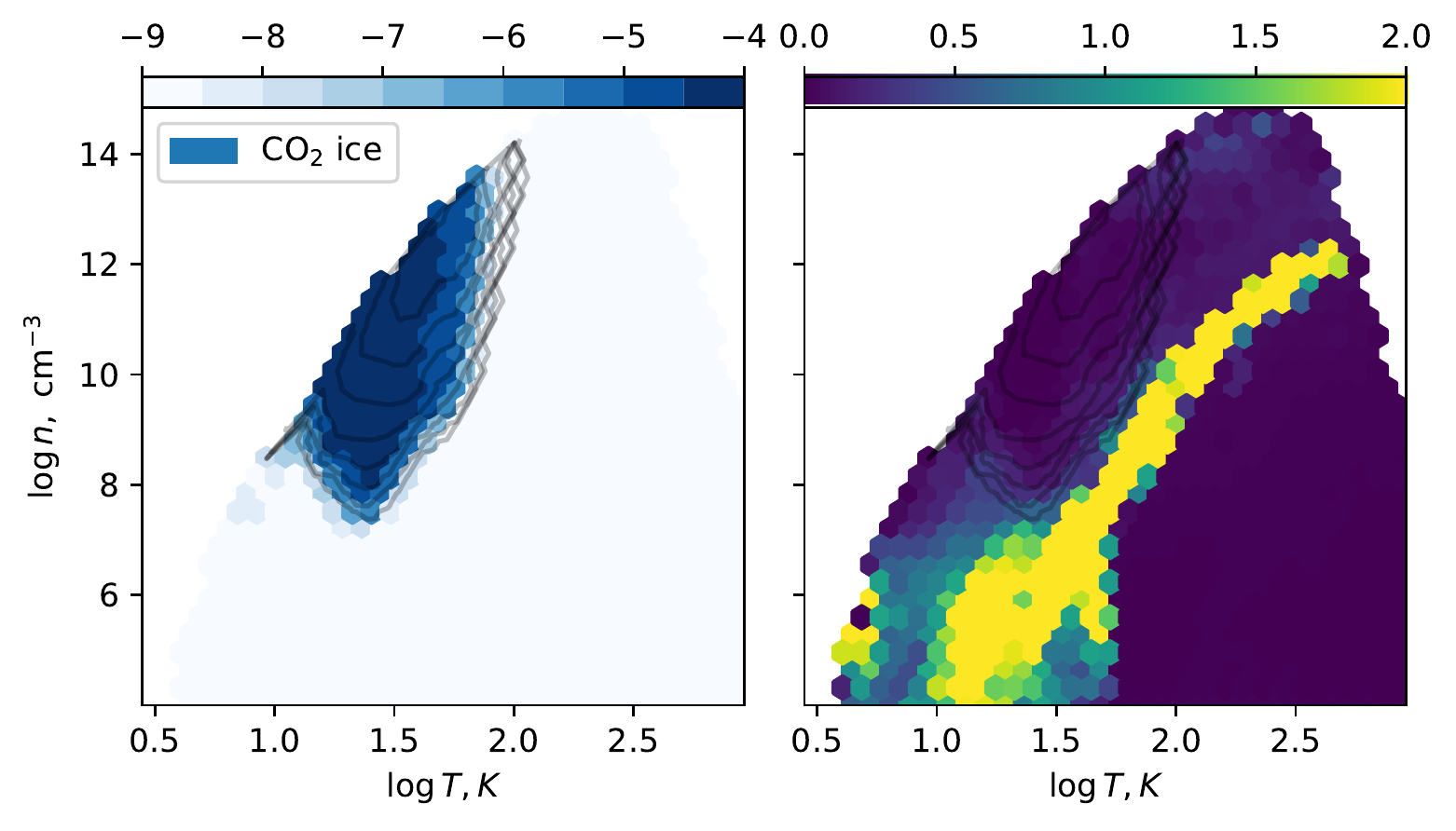}

    \includegraphics[width=0.37\linewidth]{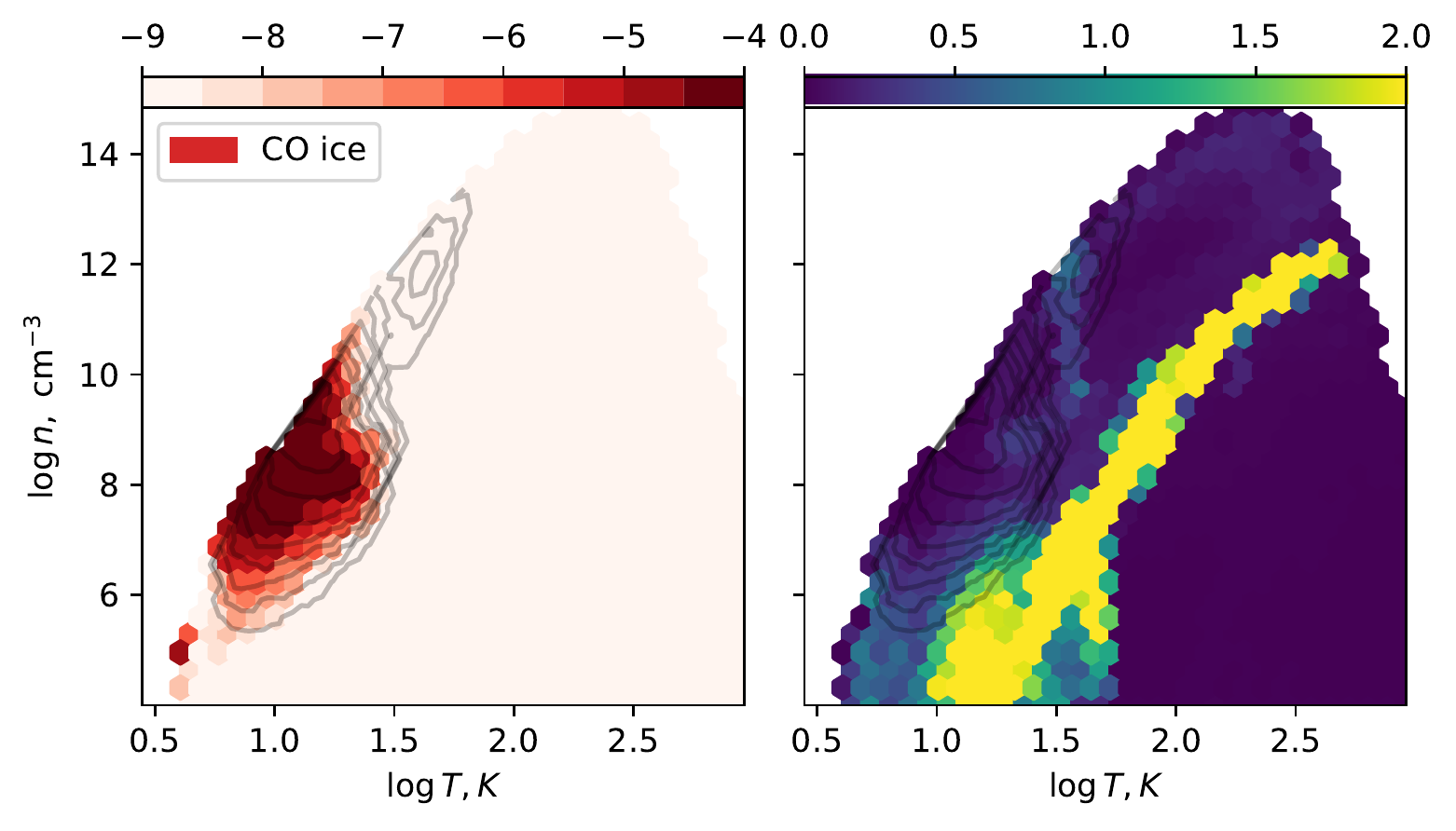}
    \includegraphics[width=0.37\linewidth]{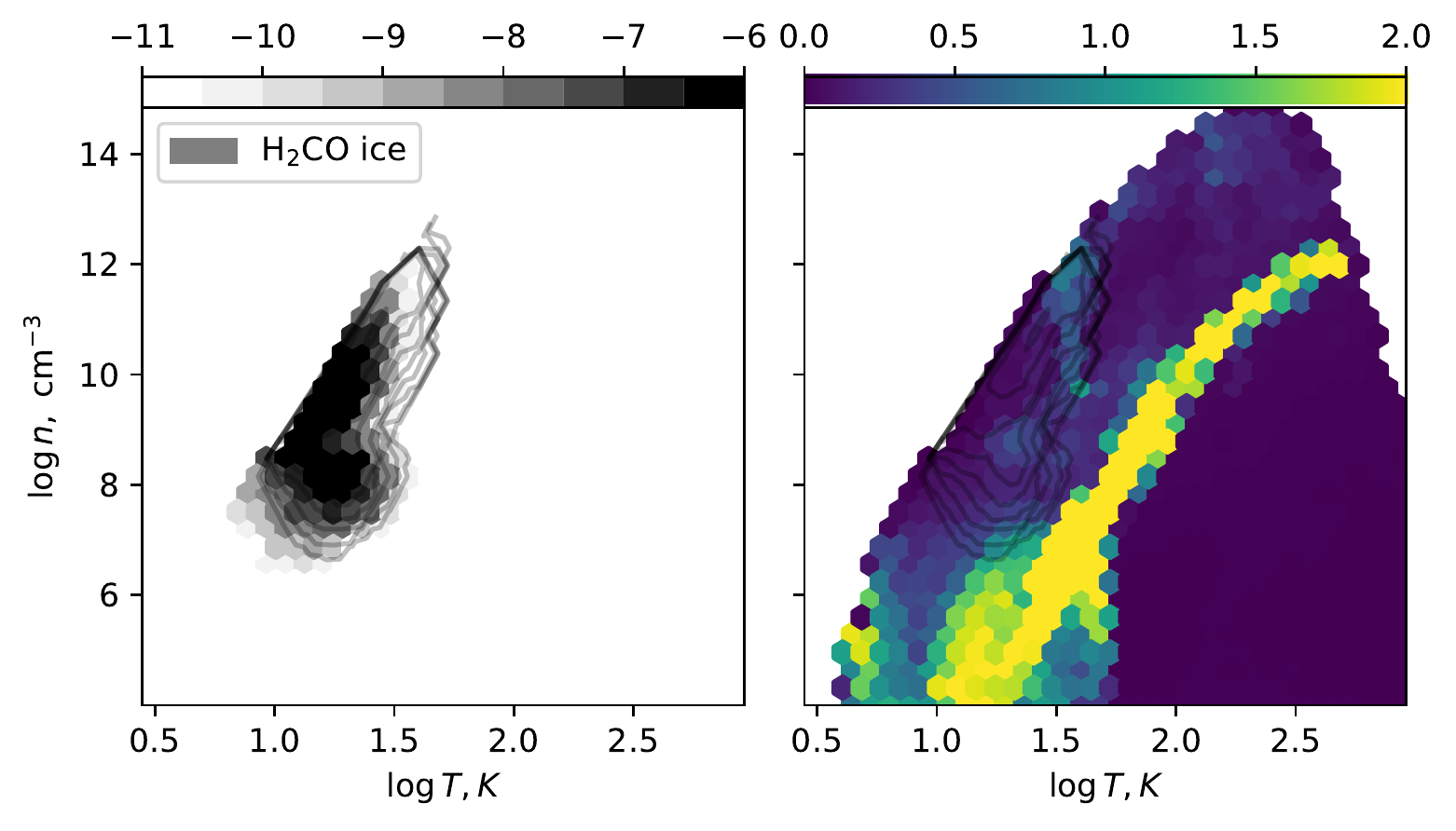}

    \includegraphics[width=0.37\linewidth]{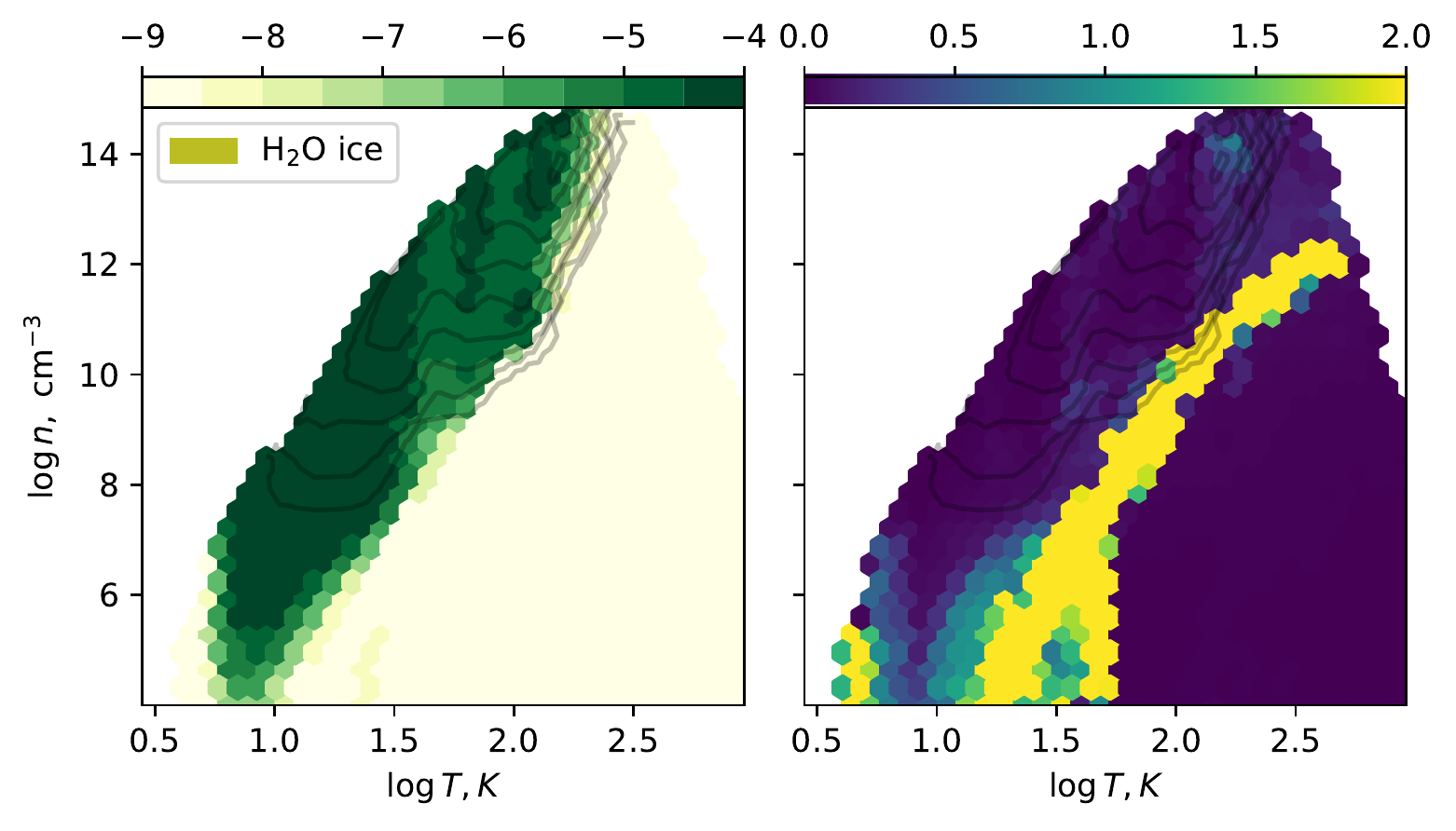}
    \includegraphics[width=0.37\linewidth]{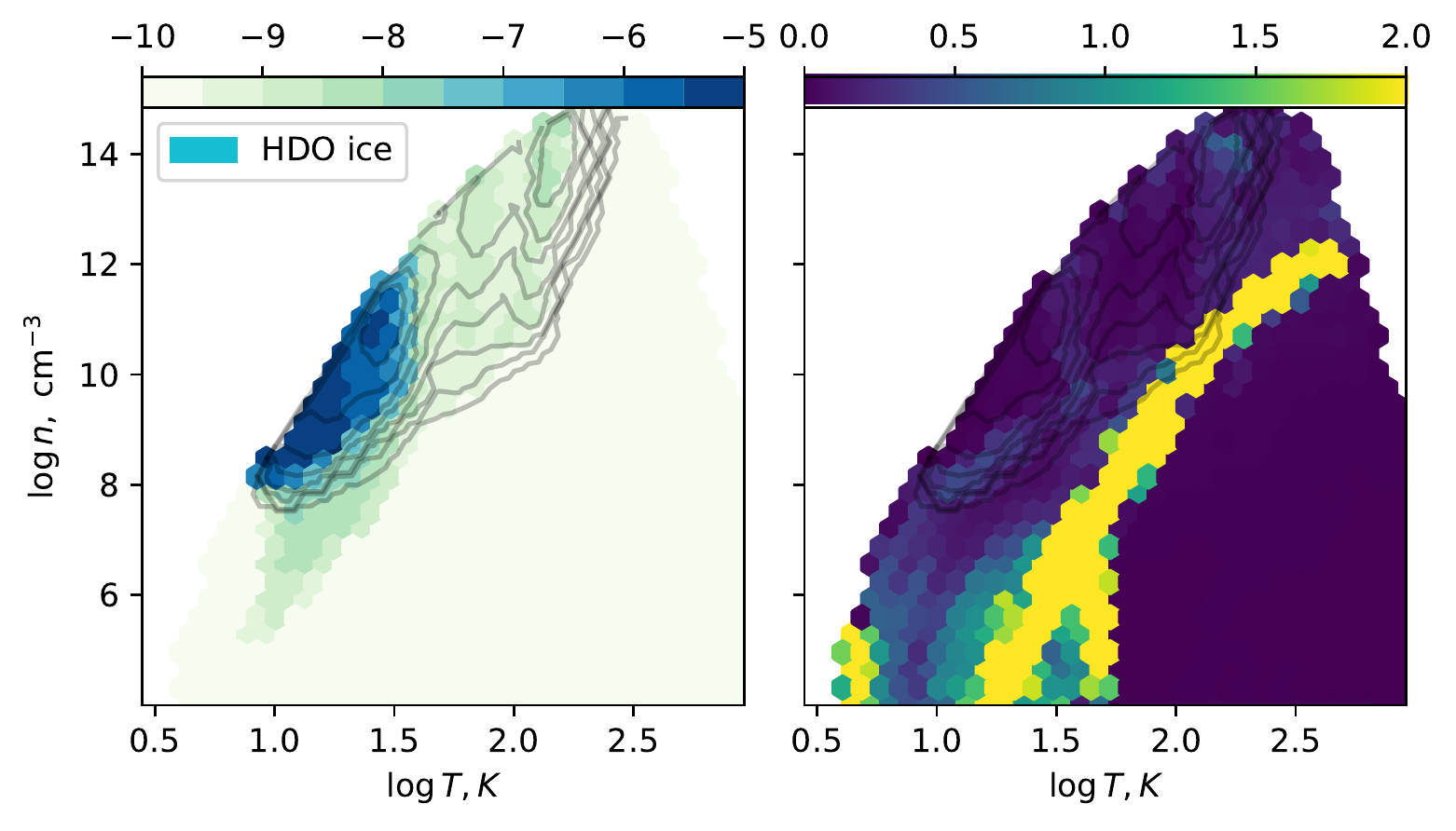}

    \includegraphics[width=0.37\linewidth]{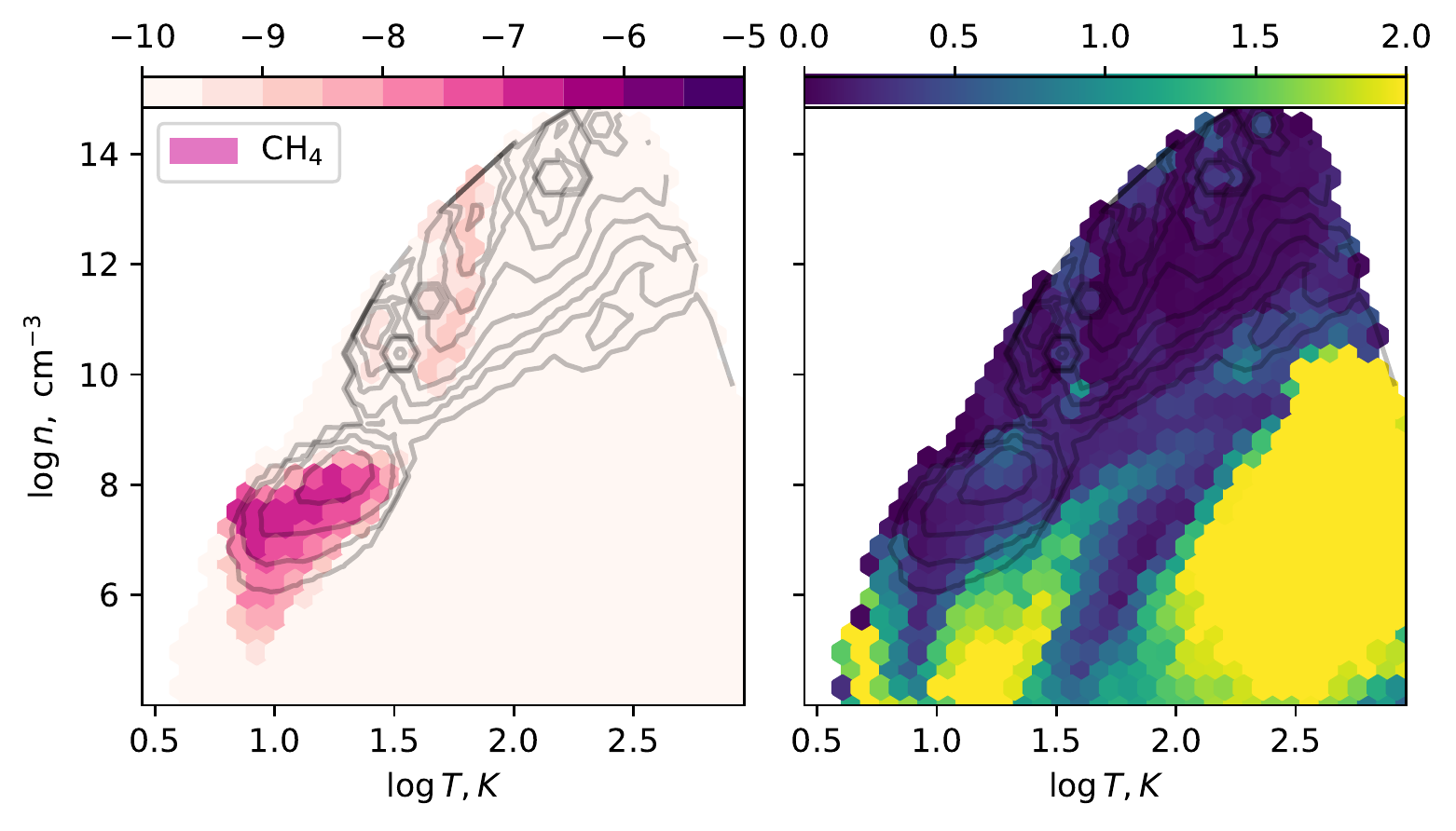}
    \includegraphics[width=0.37\linewidth]{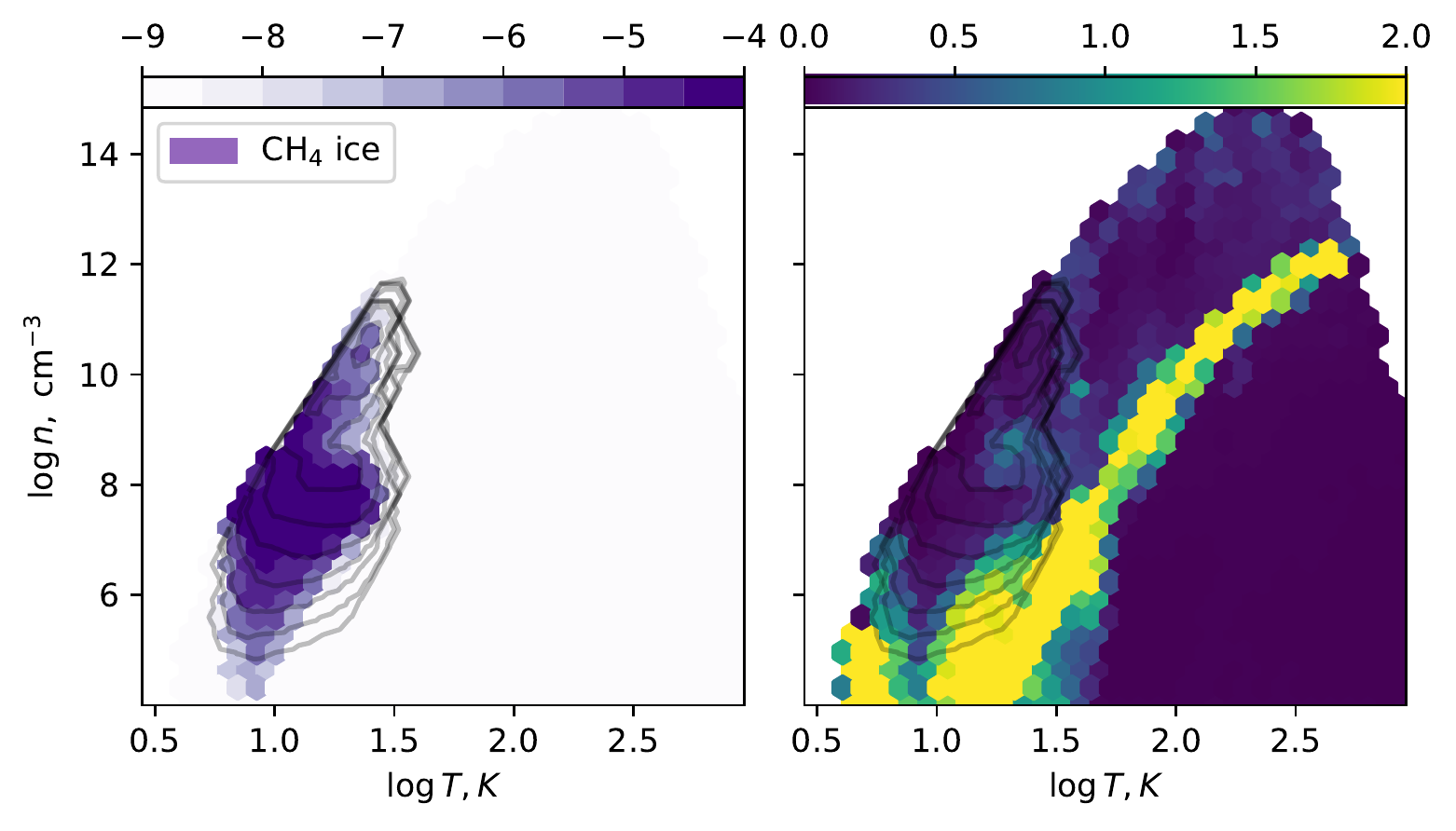}

    \includegraphics[width=0.37\linewidth]{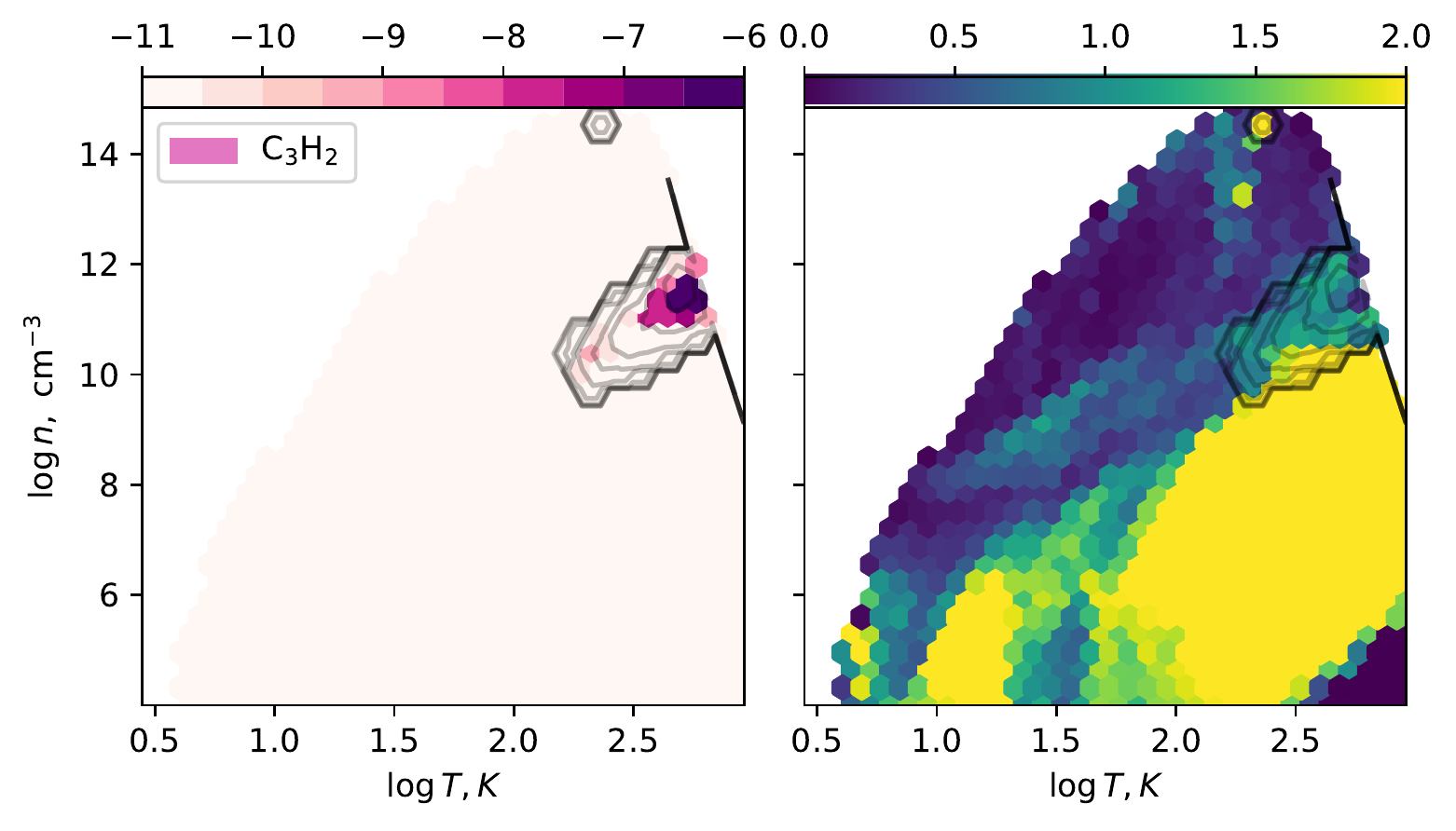}
    \includegraphics[width=0.37\linewidth]{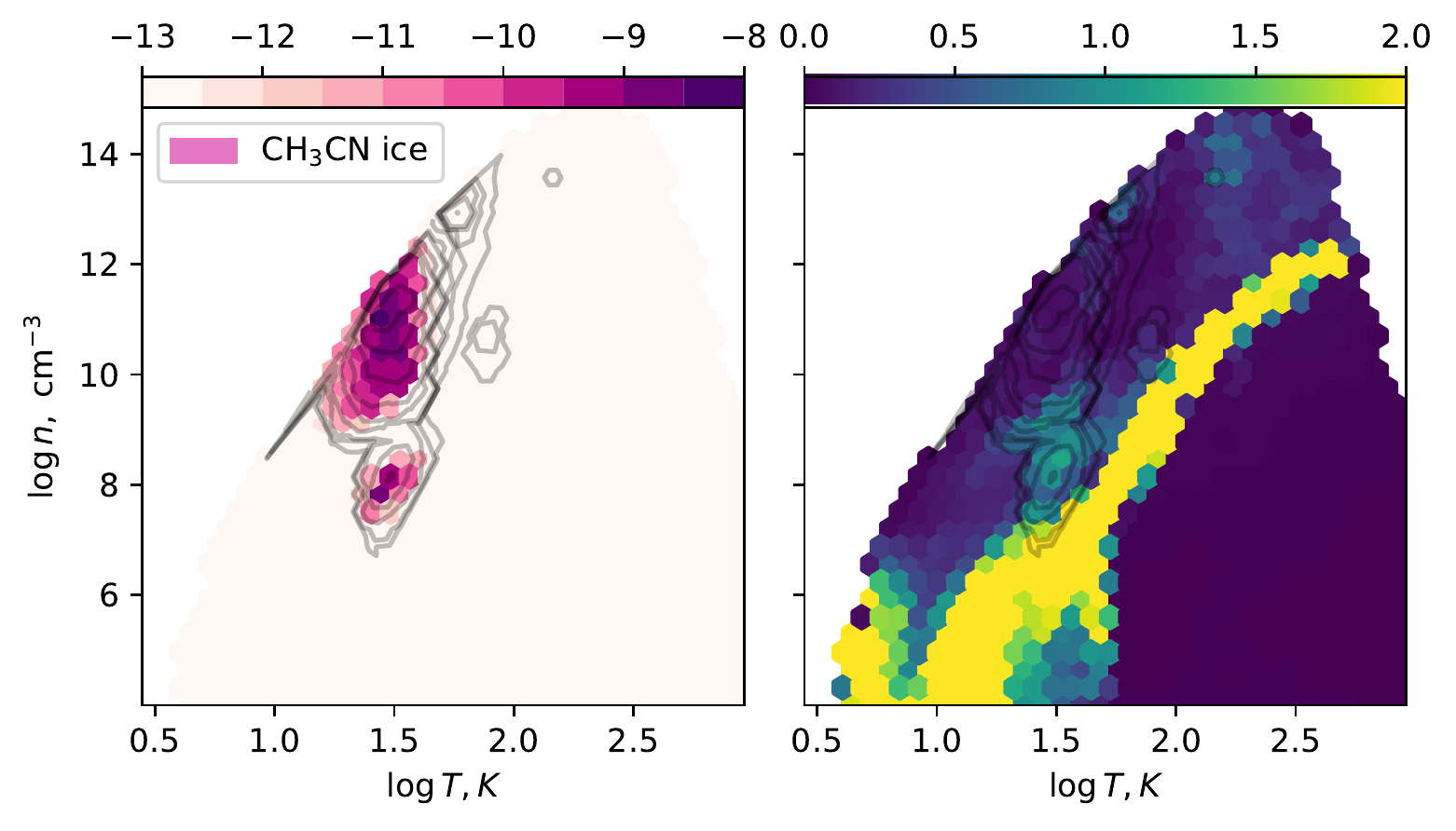}

    \caption{continued.}
    \label{fig:morespecies3}
\end{figure*}
\end{appendix}
\end{document}